\documentclass[aps,pre,twocolumn,superscriptaddress]{revtex4-2}

\usepackage{graphicx}
\usepackage{amsmath}
\usepackage{amssymb}
\usepackage{hyperref}
\usepackage{xcolor}
\usepackage{siunitx}
\usepackage{bm}

\usepackage{xr}
\begin{document}

\title{Intermittent two-phase flow in porous media:\\
insights from pore-scale direct numerical simulation}

\author{Alexandra Karabasova}
\affiliation{Department of Earth Science and Engineering, Imperial College London}

\author{Sajjad Foroughi}
\affiliation{Department of Earth Science and Engineering, Imperial College London}

\author{Martin J. Blunt}
\affiliation{Department of Earth Science and Engineering, Imperial College London}

\author{Branko Bijeljic}
\affiliation{Department of Earth Science and Engineering, Imperial College London}

\date{\today}

\begin{abstract}
Recent X-ray imaging experiments have revealed that multiphase flow through porous media involves transient fluctuations in local occupancy, even under fixed macroscopic steady-state conditions where capillary forces dominate at the pore scale. To examine how intermittency manifests at the pore scale we perform direct numerical finite volume simulations (DNS) of immiscible two-phase flow through a micro-CT-derived Bentheimer sandstone geometry at capillary numbers in the Darcy and intermittent flow regimes.
We show that intermittent disconnection and reconnection are accompanied by strongly coupled local pressure redistribution and non-wetting phase flow. 
This behaviour contrasts with the Darcy flow regime, in which the phases remain predominantly in fixed pathways.
Macroscopically the computed pressure-gradient-capillary-number relationship ($\nabla P$-Ca) recovers both the linear Darcy and the sub-linear intermittent scaling regimes consistent with previous experimental measurements. 
We show how an increase in intermittency leads to the transition from the linear to the sub-linear regime.
Using topology-aware snap-off detection, we show that the spatial extent of intermittency increases with capillary number.
Spectral, local-geometry, and network-connectivity analyses provide further evidence that the intermittent elements organise into connected conduits embedded within a stable backbone of fixed flow pathways: intermittency is a network-coupled rather than purely local process. 
This work characterises the pore-scale manifestation of intermittency as a periodic sequence of drainage and imbibition displacements triggered by local pressure fluctuations whose macroscopic consequence is to improve the overall mobility of the fluid phases.

\end{abstract}

\maketitle

\section{Introduction}
\label{sec:intro}

Multiphase flow through porous media governs processes of fundamental importance in the geosciences and engineering, from hydrocarbon recovery and geological carbon sequestration to groundwater remediation, hydrogen storage~\cite{Del22} and processes in manufactured media, including packed bed reactors, flow batteries and fuel cells~\cite{Par07, Suj06}.
Classical descriptions of multiphase flow at the continuum scale rest on the extended Darcy framework, which treats the flow as locally steady and parametrises it through relative permeability and capillary pressure functions of saturation alone~\cite{Bea72}.
Implicit in this framework is the assumption that, under constant macroscopic forcing, the pore-scale fluid configuration reaches an equilibrium state with a linear relationship between flow rate and pressure gradient.

A growing body of experimental evidence has overturned the conventional Darcy-like assumptions for multiphase flow in porous media.
Using confocal microscopy, Datta \emph{et al.}~\cite{Dat14} showed that trapped non-wetting ganglia remain unchanged at low flow rate but begin to mobilise above a threshold, consistent with a balance between viscous forcing and pore-scale capillary trapping.
Reynolds \emph{et al.}~\cite{Rey17} used time-resolved X-ray imaging to demonstrate dynamic fluid connectivity during steady-state multiphase flow in sandstone, showing that fluid-fluid interfaces within individual pores rearrange continuously during nominally steady-state coinjection.
Gao \emph{et al.}~\cite{Gao17,Gao20} subsequently quantified this phenomenon, termed \emph{intermittency}, using a combination of macroscopic pressure measurements and pore-scale X-ray imaging, establishing a direct link between pore-scale dynamics and the macroscopic multiphase Darcy law.

Based on the work described above, there is an emerging consensus that multiphase flow can be divided into three distinct flow regimes~\cite{Ber26} as a function of the capillary number,
\begin{equation}
    \mathrm{Ca} = \frac{\mu\, q_t}{\sigma},
    \label{Eq:Ca}
\end{equation}
where $q_t$ is the total Darcy flux, $\mu$ is the average dynamic viscosity of the two fluid phases, and $\sigma$ is the interfacial tension.

At the lowest $\mathrm{Ca}$ (regime I), the pressure gradient is proportional to flow rate, corresponding to a linear Darcy regime: 
\begin{equation}
    q_t = -{K\lambda_t}\nabla P 
    \label{Eq:Darcy}
\end{equation}
for a porous medium of permeability $K$. $\lambda_t$ is the total mobility that can be defined in terms of relative permeability $k_r$ as $\lambda_t = k_{r1}/\mu_1 + k_{r2}/\mu_2$ where the labels 1 and 2 indicate the two phases. In this paper, phase 1 is the wetting phase and phase 2 the non-wetting phase. This linear relationship implies either that the fluid configuration is fixed, or that, if there are regions of fluctuating occupancy, these contribute negligibly to flow, or represent a fixed intermittent fraction where locally the fluid flux is proportional to pressure gradient. 

At higher $\mathrm{Ca}$, but where capillary forces still dominate at the pore scale, dynamic fluctuations persist and the relationship between pressure gradient and flow rate becomes nonlinear, with a sub-linear power-law dependence of pressure gradient on capillary number. This is the intermittent regime II: 
\begin{equation}
   \nabla P \sim q_t^m \sim \mathrm{Ca}^{\,m}
\label{Eq:Intermittent}
\end{equation}
with $m<1$~\cite{Gao20,Zha21}. As flow rate increases, more of the pore space experiences intermittent fluctuations: the alternating occupancy of critical regions of the pore space facilitate flow, increasing the total mobility, leading to the nonlinear behaviour observed empirically, Eq.~(\ref{Eq:Intermittent}) \cite{Rey17}.

At even higher $\mathrm{Ca}$, most of the pore space experiences intermittent occupancy when viscous forces begin to dominate at the pore scale.  This leads to regime III, viscous domination or ganglion dynamics, where again a linear relationship between flow rate and pressure gradient is recovered, before inertial effects begin to have an effect.

Spurin \emph{et al.}~\cite{Spu21} showed, using both synchrotron and laboratory-based X-ray imaging, that  intermittent pores organise into system-spanning pathways that enable both phases to traverse the medium simultaneously through temporally multiplexed routes.
In parallel, several complementary descriptions have been proposed for the pore-scale physics underlying the intermittent regime.
Hansen and co-workers~\cite{hansen2018relations,roy2019flow} developed a statistical thermodynamic framework in which fluid-fluid interface motion is treated as a sequence of threshold-crossing events governed by local capillary barriers, yielding testable predictions for the nonlinear pressure-gradient scaling.
The thermodynamically constrained averaging theory (TCAT) of Hassanizadeh and Gray~\cite{Has93} provides a rigorous continuum framework in which additional state variables such as interfacial area can enter the macroscale description.
Zhang \emph{et al.}~\cite{Zha21} added an energy-dissipation perspective, interpreting intermittency as a transport mode that appears when viscous forcing repeatedly destabilises interfaces that would otherwise remain pinned by capillary resistance.

What is still missing, however, is a physically-based understanding of how intermittency manifests based on a realistic three-dimensional dataset that resolves interface topology, pressure redistribution, and flow structure simultaneously in a natural porous medium to predict and interpret intermittent behaviour and its occurrence even when capillary effects dominate at the pore scale. 
This work will focus on the transition between regimes I and II through a numerical investigation of multiphase flow through a realistic rock geometry.

Direct numerical simulation (DNS) offers a powerful complement to experiment: it resolves the full Navier-Stokes equations coupled to interface transport on a representative pore geometry, providing spatially and temporally resolved predictions of the phase, pressure and velocity fields. The strength of realistic 3D DNS is therefore not that it invalidates reduced descriptions, but that it can test and interpret them on realistic pore geometries at high resolution~\cite{Rae12,Sha18}.
It provides simultaneous access to interface motion, pore-scale pressure fields, and local flow pathways in a way that is extremely difficult to obtain experimentally and impossible in lower-dimensional idealisations alone.
However, to date no DNS study has systematically characterised intermittency and its spectral properties in such a geometry, nor directly resolved the capillary pressure dynamics at individual pore-space elements during intermittent events.

In this paper, we present DNS of immiscible two-phase flow through a Bentheimer sandstone image subvolume using a volume-of-fluid (VOF) solver. We address the following objectives:
\begin{enumerate}
\item Provide a direct pore-scale visualisation of an intermittent event, resolving the local sequence of pore-scale displacement and the surrounding pressure and velocity fields (Sec.~\ref{sec:p6_cluster}).
\item Map the network-scale organisation of fixed and intermittent flow paths, and quantify how the intermittent fraction grows with capillary number (Sec.~\ref{sec:network_pathways}).
\item Demonstrate that the DNS reproduces the macroscopic Darcy and intermittent $\nabla P$-$\mathrm{Ca}$ scaling observed experimentally (Sec.~\ref{sec:macro}).
\item Characterise the temporal organisation of the intermittent activity via spectral analysis (Sec.~\ref{sec:temporal}).
\item Connect the intermittent activity to local pore geometry and to the formation of system-spanning pathways (Sec.~\ref{sec:pore_structure}).
\end{enumerate}
Before presenting the main results, Section~\ref{sec:methods} describes the simulation methodology, geometry, and post-processing: the DNS solver, the adaptive steady-state detection algorithm, the local topology-based intermittency classification, and the pressure-gradient extraction used for comparison with macroscopic experiments.

\section{Methods}
\label{sec:methods}

This section describes the numerical and post-processing framework used. To interpret the results, a topologically equivalent network was extracted from the pore-space image used for the simulations: using the maximal ball method pores (wide regions of the pore space) and throats (constrictions between pores) were identified~\cite{dong2009pore}.
The selected $100^3$-voxel subvolume ($\SI{358}{\micro\metre} \times \SI{358}{\micro\metre} \times \SI{358}{\micro\metre}$) contains a relatively small but connected pore network of 28~pores and 39~throats, with a mean coordination number of~3.3, see the Supplementary Material S1.

For flow simulation, the segmented pore space was converted into a finite-volume mesh which mapped each pore voxel directly to a hexahedral control volume and yielded a mesh with 157\,784 cells.
Three iterations of boundary and internal smoothing (relaxation factor 0.1) were then applied to improve mesh quality without materially altering the pore geometry.
The resulting mesh had a maximum non-orthogonality of $28.4^\circ$ and a maximum aspect ratio of~2.5, which are comfortably within the range typically regarded as acceptable for stable and accurate finite-volume calculations in complex porous geometries~\cite{FoamExtend41}.

\subsection{Governing equations and numerical method}

We solved the incompressible, immiscible two-phase Navier-Stokes equations using the volume-of-fluid (VOF) method as implemented in the \texttt{poreFoam2f} solver~\cite{Rae12,Sha18, Imp26pf}, an extension of the \texttt{icoNSFoam} solver within the OpenFOAM framework (specifically foam-extend 4.1) adapted for porous media applications.
The method tracks the phase distribution through a scalar field $\alpha(\bm{x},t)$, where $\alpha = 1$ denotes the wetting phase and $\alpha = 0$ the non-wetting phase.
The governing equations are the continuity, momentum, and VOF transport equations:
\begin{equation}
  \nabla \cdot \bm{u} = 0, \label{eq:continuity}
\end{equation}
\begin{multline}
  \frac{\partial (\rho\,\bm{u})}{\partial t}
    + \nabla\cdot(\rho\,\bm{u}\bm{u})
  = -\nabla p \\
    + \nabla\cdot\bigl[\mu\bigl(\nabla\bm{u}
    +\nabla\bm{u}^T\bigr)\bigr]
    + \bm{f}_\sigma, \label{eq:momentum}
\end{multline}
\begin{equation}
  \frac{\partial \alpha}{\partial t}
    + \nabla\cdot(\alpha\,\bm{u})
    + \nabla\cdot\bigl[\alpha(1{-}\alpha)\,\bm{u}_c\bigr]
  = 0, \label{eq:vof}
\end{equation}
where $\bm{u}$ is the velocity field, $p$ the pressure, and $\bm{f}_\sigma = \sigma\kappa\nabla\alpha$ is the interfacial tension force modelled via the continuum-force formulation (surfaceForceModel = CCF), within the broader continuum-surface-force framework of Brackbill~\emph{et al.}~\cite{Bra92}.
The compression velocity $\bm{u}_c$ maintains a sharp interface.
Fluid properties are interpolated as $\rho = \alpha\rho_w + (1{-}\alpha)\rho_o$ and likewise for $\mu$.

Both solvers used a semi-implicit time integration scheme, treating convective terms explicitly and diffusive terms implicitly to ensure numerical stability. Because the VOF interface is diffuse over about two voxels, additional uncertainty arises only when quantities are derived across the interface itself.
For the local pore-cluster capillary-pressure estimates used later, the narrowest throat was T20, with inscribed radius $r_\mathrm{min}=\SI{9.86}{\micro\metre}$.
This gives a geometric capillary-pressure scale of order
$\Delta p_c \sim 2\sigma/r_\mathrm{min} \approx 6.1\times 10^3\mathrm{Pa}$
for $\sigma=\SI{0.03}{N\per m}$.
This remains below the characteristic dynamic-pressure fluctuation amplitude observed of order $\SI{1.0e4}{Pa}$.

\subsection{Fluid properties and boundary conditions}

Both phases 1 (wetting) and 2 (non-wetting) had equal density ($\rho = \SI{1000}{kg\per m^3}$) and equal dynamic viscosity ($\mu = \SI{e-3}{Pa.s}$), giving a viscosity ratio $M = \mu_2/\mu_1 = 1$.
The interfacial tension was $\sigma = \SI{0.03}{N\per m}$, and the contact angle was $\theta = 15^\circ$ measured through phase 1.

The phase indicator was defined such that $\alpha_1 = 1$ denotes the wetting phase and $\alpha_1 = 0$ the non-wetting phase.
Wettability was imposed on the grain surfaces through a contact-angle boundary condition for $\alpha_1$, while the outlet and external non-solid faces used zero-gradient conditions and the inlet used a fixed-value phase distribution.
At the inlet, both wetting and non-wetting phase flow was imposed with equal prescribed phase flow rates, giving a wetting fractional flow of $f_1 = 0.5$.
At the outlet, a zero-gradient condition was imposed during outflow and zero velocity was prescribed if local reverse flow occurred.
The pressure fields used fixed-flux and relaxed mean-value boundary conditions to maintain incompressibility and stable interface evolution.

Simulations were parallelised by domain decomposition and run on Imperial College's CX3 Phase~2 high-performance computing cluster using a single 45-core node per job. 
We performed simulations at multiple capillary numbers (Eq.~\ref{Eq:Ca}) by varying the inlet velocity, $\mathrm{Ca} \approx 9.14 \times 10^{-7} \text{ to } 2.46 \times 10^{-5}$ to span the Darcy and Intermittent regimes.
The full set of simulated cases is listed in Table~\ref{tab:cases}.

Each case is classified as Darcy or Intermittent using two independent criteria. 
The first is the analytical criterion of Zhang \emph{et al.}~\cite{Zha21}, by which the flow is intermittent when $\mathrm{Ca}/Y^i > (1-f_1)^2$, with $Y^i = (\mu_{2}/\mu_1)\,K r^2/(\phi_{\mathrm{eff}} l^4)$. $r$ is a characteristic throat radius and $l$ is an average distance between pores. 
We use the flow-accessible porosity $\phi_{\mathrm{eff}} = 0.158$ here for consistency with the operational definition of $K$, which is computed from a single-phase flow simulation on the same geometry; the geometric porosity $\phi=0.215$ is retained separately for the conventional macroscopic Darcy-flow normalisation $K/(\mu\phi)$ used in Sec.~\ref{sec:macro}. 
For our Bentheimer geometry this gives $Y^i = 1.38\times10^{-5}$, equivalent to a threshold $\mathrm{Ca} \approx 3.4\times10^{-6}$ at $f_w = 0.5$. 
The second is the experimental dataset of Gao \emph{et al.}~\cite{Gao20}, who report a regime transition at $\mathrm{Ca} \approx 10^{-5}$. The two criteria agree on all of our cases except 100w100 ($\mathrm{Ca}=6.82\times10^{-6}$), which sits just above Zhang's threshold but below Gao's transition; we therefore label it as a transition case and exclude it from the regression fits on both regimes.
This classification is confirmed by the macroscopic $\nabla P$ - Ca scaling reported in Sec.~\ref{sec:macro} (Fig.~\ref{fig:gradP_Ca}).

\begin{table}[b]
\caption{Summary of the simulation cases. The listed $U$ values denote the nominal Darcy flux assigned to each phase in the equal-fraction inlet condition. Ca values are computed from the time-averaged inlet flux $\langle Q_\mathrm{in}\rangle$ over the steady-state window, since the prescribed inlet velocity does not fully translate to the realised flow rate at high Uo. The capillary number is defined as in Eq.~(\ref{Eq:Ca}); the regime classification is described in the main text above.}

\label{tab:cases}
\renewcommand{\arraystretch}{1.15}
\begin{ruledtabular}
\begin{tabular}{lccc}
Case & $U\;(\si{\micro m/s})$ & Ca & Regime \\
\colrule
10w10   & 10  & $9.14 \times 10^{-7}$ & Darcy\\
50w50   & 50  & $3.08 \times 10^{-6}$ & Darcy\\
100w100 & 100 & $6.82 \times 10^{-6}$ & Transition \\
200w200 & 200 & $1.30 \times 10^{-5}$ & Intermittent \\
300w300 & 300 & $1.62 \times 10^{-5}$ & Intermittent \\
400w400 & 400 & $2.01 \times 10^{-5}$ & Intermittent \\
500w500 & 500 & $2.46 \times 10^{-5}$ & Intermittent \\
\end{tabular}
\end{ruledtabular}
\end{table}

The wall-clock runtimes ranged from \SI{314}{h} (\SI{14139}{core.h}) for the highest-$\mathrm{Ca}$ case to \SI{1239}{h} (\SI{55737}{core.h}) for the longest single run, 50w50; the corresponding simulated physical times reached up to \SI{4.40}{s} (Supplementary Material S2). 
These computational costs restricted the present DNS study to a $100^3$-voxel subvolume despite the REV analysis (can be found in the Supplementary Material S1) indicating that geometric representativity would require a larger domain.

\subsection{Classification of intermittency}
\label{sec:classification}

All intermittency analyses are performed within the validated steady-state (SS) window of each simulation, identified from the domain-averaged wetting-phase saturation $\langle S_1 \rangle(t)$. The SS-window detection algorithm and the per-case windows are summarised in the Supplementary Material S3 (Table S2).

\subsubsection{Pore-scale displacement events}

Intermittency involves a repeated sequence of pore-scale displacement events.
A \emph{Haines jump} is a drainage event in which the non-wetting phase rapidly invades wider regions of the pore space once a capillary barrier is overcome at a throat~\cite{haines1930studies}. This can reconnect the non-wetting phase.
\emph{Snap-off} is an imbibition event in which the wetting phase swells through corners or films and disconnects the non-wetting phase in a throat~\cite{Pic66}.
\emph{Cooperative pore filling} is also an imbibition event that occurs when neighbouring menisci advance in a correlated way so that invasion of a pore becomes favourable only after adjacent throats have already filled~\cite{Len83}.
In this paper, intermittency is interpreted as a recurrent sequence of such drainage and imbibition events, leading to repeated disconnection and re-connection of the phases.

\subsubsection{Element-based phase-connectivity classifier}
\label{sec:element_classifier}

For each pore or throat and each saved timestep within the steady-state window, a connectivity test determines whether the non-wetting phase spans an element (pore or throat) from one side to the other.
For throats this means that a connected non-wetting cluster links the two pore interfaces; for pores it means that the non-wetting phase connects between at least two throat interfaces.
This yields a boolean per-element time series $s_j(t)\in\{0,1\}$ describing whether the phase is connected.
An element is labelled \emph{intermittent} if its signal $s_j(t)$ completes at least two full switching cycles over the steady-state window, otherwise it is labelled \emph{fixed}.
This provides a clear and regularly sampled time series for each pore or throat, which is ideal for the network-classification maps (Sec.~\ref{sec:network_pathways}), the spectral analysis (Sec.~\ref{sec:temporal}), and the pathway analysis (Sec.~\ref{sec:pore_structure}).

However, the whole pore or throat is labelled as switching even if the change is confined to a small internal sub-region.
For these reasons, we construct a more robust event-based classifier to operate on the voxel-level phase field.

\subsubsection{Fixed-site event-based phase-connectivity classifier}

To capture intermittency at sub-element resolution we use a voxel-level topology-change detector that operates on pairs of consecutive timesteps within the SS window. 
At each timestep the non-wetting phase is identified as occupying any voxels with $\alpha \leq 0.5$. 
The non-wetting phase is collected into clusters (ganglia) using 6-face connectivity. 
Any local topology change is classified as either a \emph{disconnection} (one non-wetting cluster becomes two) or a \emph{reconnection} (two non-wetting clusters become one). 
In the classical-event language introduced above, a disconnection is the topological signature of snap-off and a reconnection that of a Haines jump. 
A voxel is classified as intermittent only if it undergoes at least three full disconnection-reconnection cycles over the SS window, with each half-phase persisting for at least \SI{1}{ms} to filter out diffuse-interface noise.

We refer to this as the \emph{fixed-site} event-based phase-connectivity classifier, because recurrence must be observed at the same voxel; it is the primary spatial intermittency metric for the cross-case quantitative analysis. 
The detailed algorithm steps and parameter choices are described in the Supplementary Material S4.

\subsection{Spectral analysis}
\label{sec:spectral_method}

Spectral analysis was performed with the \emph{element-based phase-connectivity classifier}. As defined above in Sec.~\ref{sec:element_classifier}, this yields one uniformly sampled scalar signal $s_j(t)\in\{0,1\}$ for each pore or throat element, indicating whether the non-wetting phase spans that same physical object at each saved timestep.
The advantage is therefore that it has fixed spatial support, fixed interpretation through time, regular sampling, and a clean per-element state change associated with connectivity.

For each element, the mean is removed from $s_j(t)$ and the power spectral density (PSD), i.e.\ the distribution of signal variance across frequency, is estimated using Welch averaging~\cite{Wel67}, in which the signal is divided into overlapping segments and the per-segment periodograms are averaged to reduce variance.
The dominant frequency $f_\mathrm{dom}$ is then defined as the non-zero frequency at which the PSD reaches its maximum.

To compare cases fairly, all spectral statistics are computed over the same physical duration, using the last \SI{0.75}{s} of the validated steady-state window for each case.
With this window length, the per-case Welch frequency resolution is in the range \SIrange{4}{6}{Hz}, so peaks at the low-frequency end are quantised at a similar scale.
The element-based method also inherits the pore/throat labelling used during pore-network extraction.

\subsection{Pathway analysis}
\label{sec:pathway_method}

Connected components of intermittent, wetting phase-occupied, and non-wetting phase-occupied pores and throats are identified from the pore-network graph.
Here too, the relevant signal is the element-based connectivity state of pores and throats (Sec.~\ref{sec:element_classifier}), rather than the fixed-site per-voxel event count, because pathway analysis requires a single, well-defined intermittent state for each pore or throat in the network.
The connected-component decomposition is used at the network scale to identify intermittent conduits (Sec.~\ref{sec:network_pathways}); two complementary inlet-to-outlet route definitions are used (Sec.~\ref{sec:pore_structure}).

The \emph{hydraulic} pathway is the inlet-to-outlet sequence of pores carrying the largest harmonic-mean flux, identified by Dijkstra's algorithm on a graph weighted by inverse throat flow rate $1/Q$ \cite{Dij59}.
The harmonic mean is taken over the full steady-state window, which suppresses the influence of outlier timesteps with anomalously high local flow rates.

The \emph{geometric shortest} pathway is the inlet-to-outlet path of minimum total length (sum of inscribed throat lengths plus pore-to-throat segment lengths) among all paths sharing the minimum throat-hop count through the static pore-network graph.
It depends only on the rock geometry and is therefore Ca-independent by construction.

\subsection{Macroscopic pressure gradient}
\label{sec:gradP_method}

We computed the pressure gradient $\nabla P$ from each simulation and compared with the experimental data of Gao \emph{et al.}~\cite{Gao20} and Zhang \emph{et al.}~\cite{Zha21}. 
We extracted the pressure gradient from a sub-box of streamwise extent $L_{\mathrm{sub}}$ centred about the mid-domain ($x/L_{\mathrm{domain}} = 0.5$), by linear regression of the cross-section-averaged $\langle p_d\rangle$ against $x$.
We adopted $L_{\mathrm{sub}}/L_{\mathrm{domain}} = 0.50$ throughout, excluding $25\%$ from each end. 
The rationale for this approach is described in the Supplementary Material S5. The slope is then taken from a linear regression of the trimmed-mean profile over the analysis window.
The resulting gradient is denoted $\nabla P_{\mathrm{sub}}$ throughout.

For comparison across datasets, the plotted pressure gradients were normalised by $K/(\mu\phi)$ using, for the DNS, $K=\SI{1.46347e-12}{m^2}$, $\mu=\SI{1.0e-3}{Pa.s}$, and $\phi = 0.215$, the geometric porosity of the simulated subvolume measured directly from the segmented pore-space voxel count, chosen to match the convention used by Gao \emph{et al.}~\cite{Gao20} ($\phi = 0.20$) and Zhang \emph{et al.}~\cite{Zha21} ($\phi = 0.20$). 
The two reference experiments use $K = \SI{1.9e-12}{m^2}$, $\mu = \SI{0.83e-3}{Pa.s}$ (Gao et al.) and $K = \SI{1.85e-12}{m^2}$, $\mu = \SI{0.83e-3}{Pa.s}$ (Zhang et al.).

For the macroscopic comparison, the plotted vertical error bars are computed by combining three contributions.

(i)~\emph{Averaging uncertainty}: the bootstrap standard error of the trimmed mean, estimated from 1000 resamples over the SS window.

(ii)~\emph{Steady-state detection uncertainty}: sensitivity to residual saturation drift, estimated as $|d\Delta P/dS_1| \times \mathrm{MAD}(S_1)$, where $\mathrm{MAD}$ is the median absolute deviation of the wetting phase saturation within the SS window.

(iii)~\emph{Upscaling uncertainty}: for the centred sub-box panel, the per-snapshot scatter of the linear-regression slope at $L_{\mathrm{sub}}/L = 0.50$ is converted to a standard error of the mean as $\sigma/\sqrt{N}$ with $N=100$.

Horizontal error bars in Ca were obtained separately from the variability of the inlet flux over the same SS window, using a block-averaged standard error of $Q_{\mathrm{in}}/A_{\mathrm{Darcy}}$ and propagating it into the capillary number.

\section{Results and Discussion}
\label{sec:results}
This part presents the DNS results in five linked stages.
In Sec.~\ref{sec:p6_cluster}, a representative pore-cluster analysis uses visualisation, topology tracking, and flux-balance diagnostics to demonstrate the key features of an exemplar single intermittency event.
This is followed by Sec.~\ref{sec:network_pathways}, in which the network-phase maps identify where fixed flow paths and unpinned intermittent conduits appear at the pore-network scale, and reports the dependence of the intermittent pore-space fraction on Ca. 
In Sec.~\ref{sec:macro}, the macroscopic pressure-gradient of the simulations is compared with the experimental Darcy and intermittent $\nabla P$-Ca behaviour.
Sec.~\ref{sec:temporal} and Sec.~\ref{sec:pore_structure} examine the spectral, local-geometry, and pathway organisation of the intermittent processes.

\subsection{Pore-scale manifestation of intermittency: a representative pore cluster}
\label{sec:p6_cluster}

\begin{figure*}[!t]
  \centering
  \includegraphics[width=\textwidth]{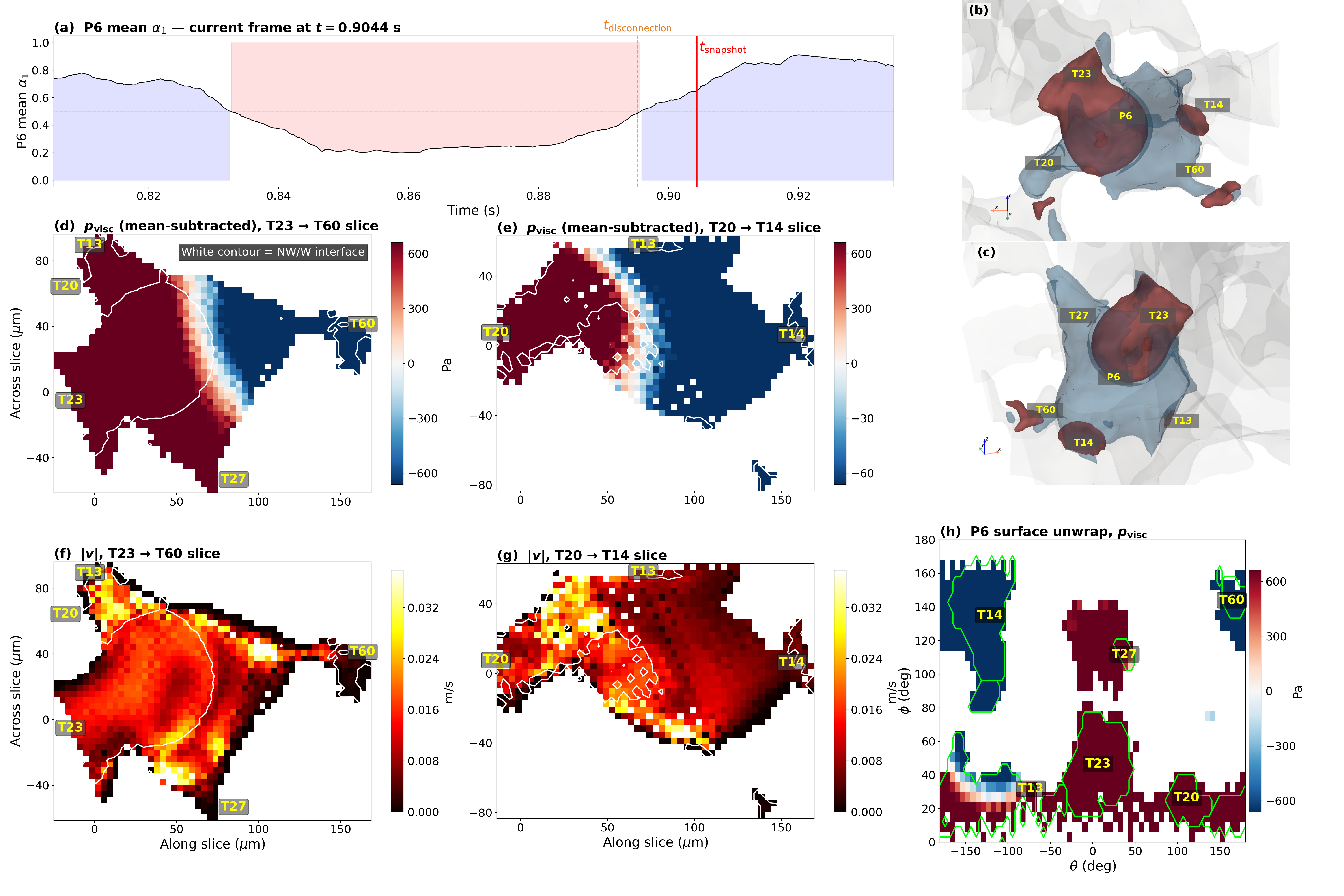}
  \caption{A representative pore-cluster intermittency event in the P6 cluster (Sec.~\ref{sec:p6_cluster}) for the transitional $\mathrm{Ca}\approx 6.82\times10^{-6}$ case, at a snapshot taken shortly after a disconnection of the non-wetting phase. (a)~Time history of the spatially-averaged phase indicator $\alpha_1$ in P6; the disconnection time is marked with the dashed orange line, and the snapshot time is labelled and marked with the vertical red line. The remaining panels are taken at that time. (b),(c)~Mean-subtracted viscous-pressure ($p_\text{visc}$) heatmaps on two orthogonal slices through the cluster, oriented T23$\rightarrow$T60 (b) and T20$\rightarrow$T14 (c); the white contour is the fluid-fluid interface and the labelled throats indicate the local network context. (d),(e)~3D pore-cluster geometry and phase configuration, viewed from two angles. (f),(g)~Velocity magnitude on the same two slices, showing the redistribution of local throughflow as the event approaches. (h)~Unwrapped P6 surface viscous-pressure field $p_\text{visc}$, summarising the viscous-pressure heterogeneity around the pore boundary.}
  \label{fig:p6_overview}
\end{figure*}

Intermittency at the pore scale consists of repeated, reversible multiphase displacement events at the narrowest constriction of a connected pore cluster, accompanied by pressure redistribution. 
We resolve this directly in physical space, with high temporal (sub-\si{\micro\second}) and spatial (\si{\micro\metre}) resolution, using a representative cluster centred on pore P6 for the transitional $\mathrm{Ca}\approx 6.82\times10^{-6}$ case. 
Fig.~\ref{fig:p6_overview} shows a single snapshot of the cluster shortly after a disconnection event, and the associated video provided in Appendix~\ref{sec:p6_video} resolves the entire intermittent cycle ($t = \SI{0.8154}{s}$ to $\SI{0.9246}{s}$) which encompasses the disconnection event shown in the snapshot.

In this cluster, we observe that the intermittent event is not a single process, but rather it is a composite of multiple individual physical processes. 
The time-resolved sequence in the video shows that neighbouring pores connected to P6 lose non-wetting occupancy before the fragmentation event in P6, after which the non-wetting ganglion inside P6 becomes disconnected and the detached fragment is advected out of the cluster.
The cycle closes when the non-wetting phase later re-invades the connecting throats through drainage (a Haines jump) and re-establishes connectivity in P6.

Each cycle maps onto a sequence of classical pore-scale displacement events. 
(i) The coordinated wetting-phase advance into the throats bordering P6 is a cooperative pore filling event~\cite{Len83}: the menisci in multiple surrounding throats advance together.
(ii) The local topological failure itself is snap-off~\cite{Pic66}: the wetting phase swells at the narrow T60 constriction and pinches off the non-wetting phase. 
(iii) The subsequent reconnection of the cluster proceeds by a Haines jump~\cite{haines1930studies}: once the upstream pressure overcomes the local capillary barrier, the non-wetting phase rapidly re-invades and restores cluster connectivity.

The intermittent event is a network-coupled process; it is a coordinated multi-element phase rearrangement within the large connected cluster as can also be seen in the local network map presented in Fig.~\ref{fig:network_phase_map_p6_focus} in Sec.~\ref{sec:network_pathways}.

To probe the actual site of the non-wetting phase disconnection, we examine the viscous-capillary pressure balance.
In the heatmaps in Fig.~\ref{fig:p6_overview}, the viscous pressure $p_\text{visc}$ consistently opposes the net driving force set by the dominant capillary pressure. 
Therefore, despite the opposition of $p_\text{visc}$, the bulk non-wetting flow proceeds from the supply throats, through P6 and out via T23.
This is the local signature of the capillary-dominated regime: the pace and triggering of the cluster's intermittent events are set by capillary forces interacting with the local pore geometry, not by the imposed bulk flow rate.

The active phase rearrangements driven by the capillary pressure field permeate into the viscous pressure gradient response.
The $p_\text{visc}$ gradient intensifies as the non-wetting phase advances through the pore via a Haines jump mechanism, peaking at a final snap-off event during the disconnection of the non-wetting phase, and then vanishes when the cluster is fully wetting phase connected (fully non-wetting phase disconnected).

These rearrangements have a clear local kinematic and capillary signature concentrated at the moving interface itself. 
The fastest local flow in the cluster is in the wetting phase along the moving interface, not in the bulk of either phase.
A transient $p_\text{visc}$ gradient maximum also appears in the slice through P6 near T20 at $t \approx \SI{0.9044}{s}$: in the time-resolved interface, this corresponds to a corner of the meniscus anchored to the P6 grain wall, with the gradient maximum persisting throughout the anchoring and disappearing the moment the ganglion detaches. 
Local capillary anchoring at a single grain corner therefore produces a $p_\text{visc}$ signature distinct from the broader cluster-scale gradient.

\begin{figure*}[t]
  \centering
  \includegraphics[width=0.96\textwidth]{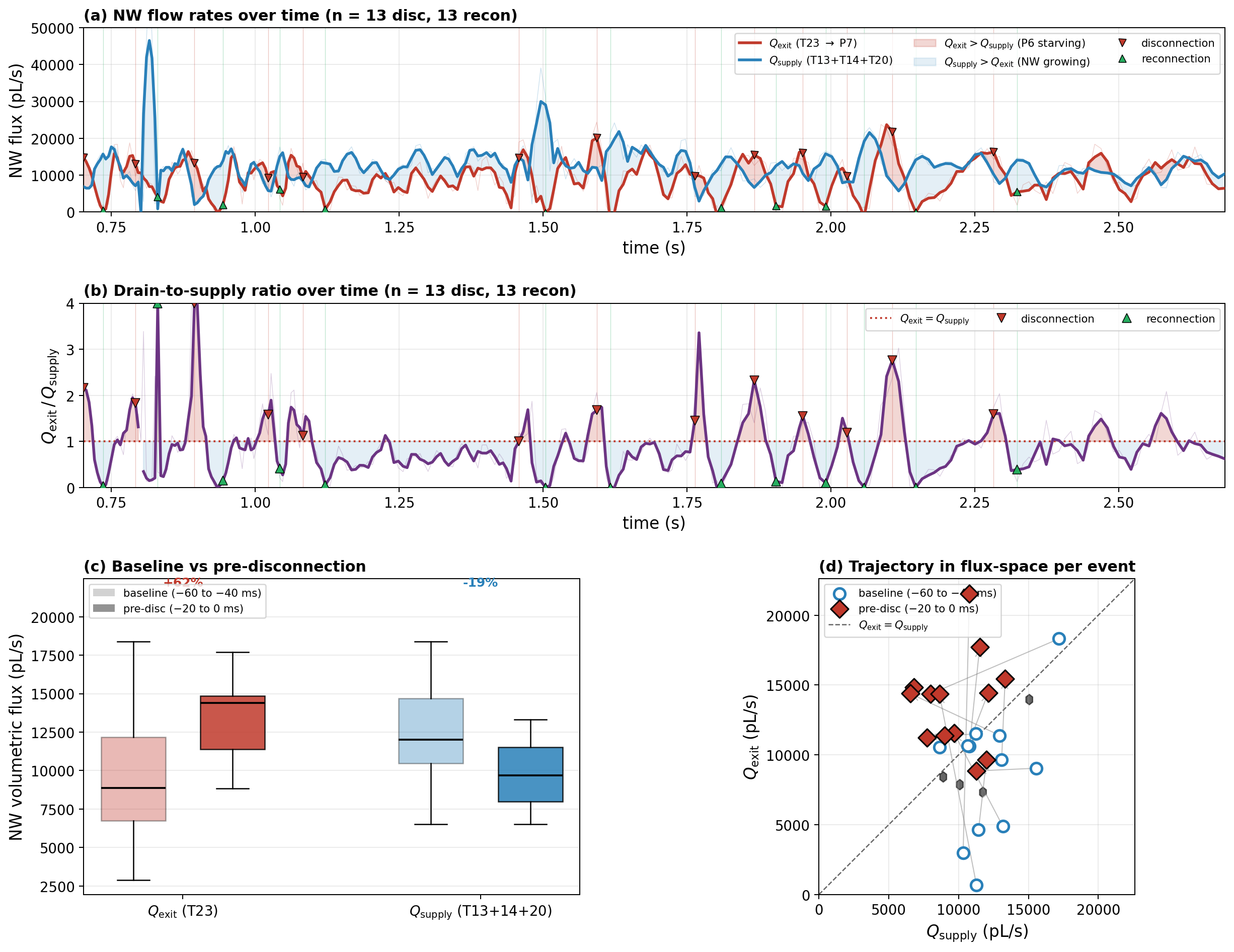}
  \caption{Flux-balance analysis for the representative P6 disconnection sequence shown in Fig.~\ref{fig:p6_overview} at $\mathrm{Ca}\approx 6.82\times10^{-6}$. (a) Full time series of non-wetting exit flux through T23 and combined non-wetting supply through T13+T14+T20, with disconnection and re-connection times marked. (b) Full-time drain-to-supply ratio, showing excursions across $Q_{\mathrm{exit}}/Q_{\mathrm{supply}}=1$ before disconnection. (c) Event-by-event windowed comparison of exit and supply fluxes: for each detected disconnection, the fluxes are averaged over a baseline interval well before the event and over a pre-disconnection interval immediately preceding it, and the box plots show the resulting distributions. The exit flux increases whereas the supply flux decreases. (d) Per-event trajectories in flux space, showing migration from a supply-dominated state to a drainage-dominated state before disconnection.}
  \label{fig:p6_flux_balance}
\end{figure*}

While these local pressure and velocity features describe the cluster's instantaneous state, the timing of disconnection itself is set by a network-scale flux dynamic. 
Snap-off at the narrow supply throats is consistently preceded by a transition from a supply-dominated to a drain-dominated flux state in the cluster.
This is a precursor that is universal across the non-wetting phase disconnection-event ensemble and cannot be captured by any local capillary-number threshold.
Fig.~\ref{fig:p6_flux_balance} compares the non-wetting exit flux through T23 with the combined non-wetting supply flux entering P6 through T13, T14, and T20 at the same transitional $\mathrm{Ca}\approx 6.82\times10^{-6}$ case as in Fig.~\ref{fig:p6_overview}.

Far from disconnection, the cluster is supply-dominated: the incoming non-wetting flux exceeds the downstream drainage rate, so the ganglion is maintained. 
This is visible in panel~(a), where the full time series of the two fluxes, together with the disconnection and re-connection times are shown.
In the approach to disconnection the exit flux rises and the supply flux weakens, so that the two curves converge and cross.
We express this explicitly in panel~(b) as a full-time drain-to-supply ratio, showing repeated excursions and, crucially, crossover of $Q_{\mathrm{exit}}/Q_{\mathrm{supply}}=1$ before disconnection events.

Across the full event ensemble, this crossover is universal. 
Panel~(c) condenses this crossover into an event-by-event windowed comparison.
For each disconnection event, the exit and supply fluxes are averaged over a baseline interval well before the event (lighter shading) and over a pre-disconnection interval immediately preceding the event (darker shading), and the box plots show the resulting distributions across the event set.
Across the 13 detected disconnection events, the exit-flux distribution shifts upward (ensemble-median $+62\%$) whereas the combined supply distribution shifts downward ($-19\%$).
Similarly, we see in panel~(d) that the corresponding per-event trajectories in flux space migrate from the quiet supply-dominated envelope to the drainage-dominated side of the $Q_{\mathrm{exit}}=Q_{\mathrm{supply}}$ diagonal before disconnection. 
The cluster therefore enters a state in which non-wetting phase is removed faster than it can be replenished, after which snap-off occurs.
In this sense, the disconnection is not explained by a purely local geometrical instability alone: a network-scale flux imbalance progressively depletes the ganglion, after which snap-off occurs at a narrow perimeter throat in the cluster.

\subsection{Network-scale pathway transition: from fixed pathways to intermittent unpinning}
\label{sec:network_pathways}
 
As Ca rises, the network transitions from a regime of pinned, permanently assigned flow pathways towards one in which a growing fraction of the pore space participates in temporally switching intermittent conduits.
To quantify this effect we use the \emph{element-based phase-connectivity classifier} (Sec.~\ref{sec:element_classifier}) for Ca in the range $\mathrm{Ca}=3.08\times10^{-6}$ to $\mathrm{Ca}=2.46\times10^{-5}$.

The network maps, Fig.~\ref{fig:network_phase_map_master}, show where the pore space behaves as a fixed permanently assigned conduit and where it is intermittent. In nearly all fixed elements both phases are present: the wetting phase typically forms layers along the grain surface and the non-wetting phase occupies the central core, the canonical two-phase configuration for a strongly water-wet medium \cite{blunt2017multiphase}.

In Fig.~\ref{fig:network_phase_map_master} the lowest flow rate ($\mathrm{Ca}=3.08\times10^{-6}$, 50w50, Table~\ref{tab:cases}) already lies close to the transition to the intermittent regime, so several elements are already intermittent. Most pores and throats keep the same classification across the analysed range, while only a small subset changes state.

\onecolumngrid
\begin{center}
  \refstepcounter{figure}\label{fig:network_phase_map_master}
  \includegraphics[width=0.96\textwidth]{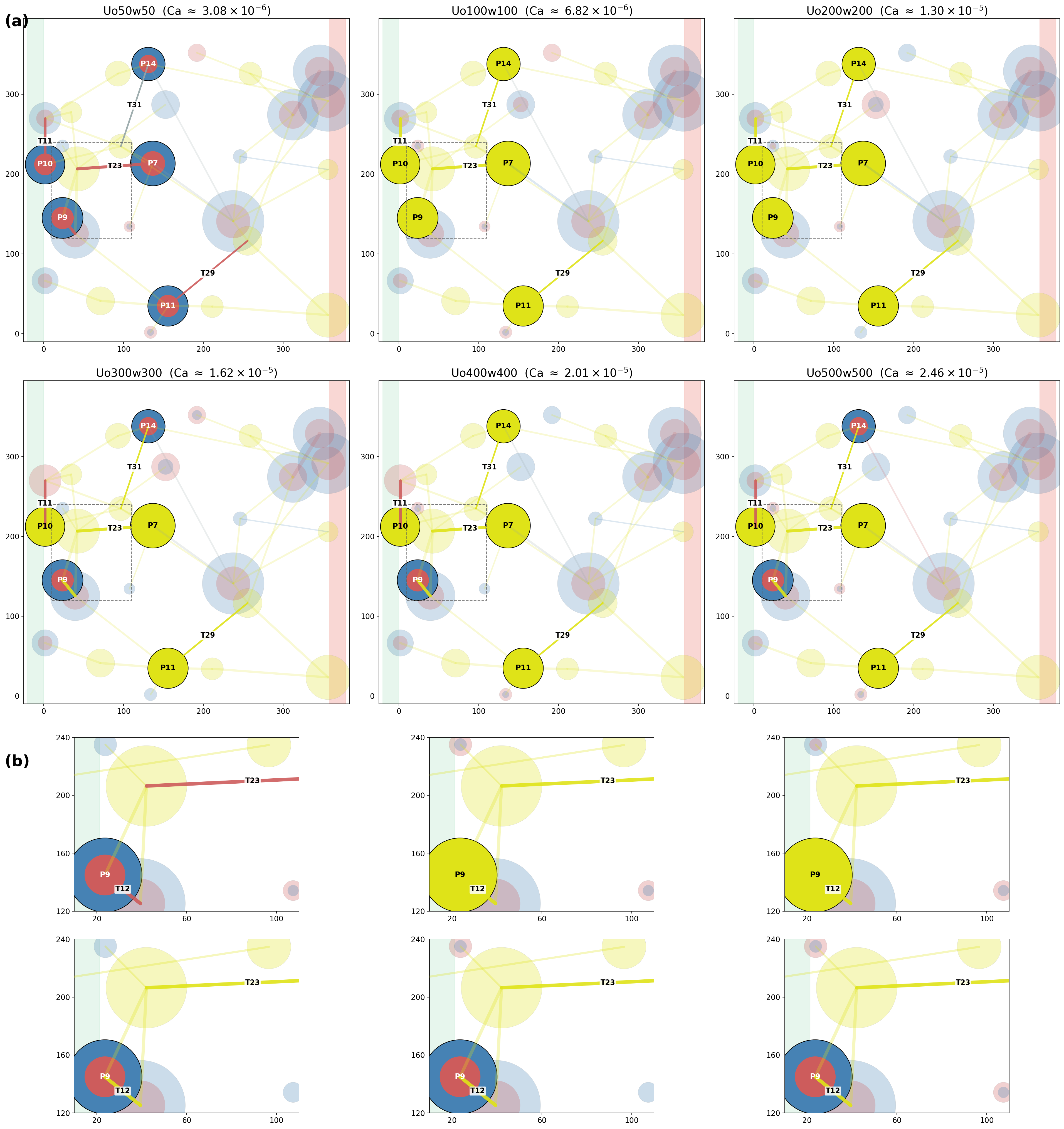}

  \vspace{0.5em}
  \parbox{0.96\textwidth}{\small\textbf{Figure \thefigure.} Pore-network phase classification as a function of capillary number, in the range $\mathrm{Ca}=3.08\times10^{-6}$ to $\mathrm{Ca}=2.46\times10^{-5}$. The network states are derived from the element-based phase-connectivity classifier: yellow elements are intermittent, blue and red indicate the wetting and non-wetting phase fixed phase occupancy respectively; faded elements do not change occupancy as $\mathrm{Ca}$ varies, whereas opaque elements change classification. (a)~Full-domain views for the six retained Ca cases (top two rows). The dashed box in each full-domain panel marks the inlet-side region shown in the zoom directly below in (b). (b)~Inlet-side zooms for the same six cases (bottom two rows), in matching column order, where the crowded local throat labels are shown explicitly.}
\end{center}
\twocolumngrid

This $\mathrm{Ca}$-sensitive subset is limited to six pores (P6, P7, P9, P10, P11, P14) and seven throats (T11, T12, T13, T14, T23, T29, T31).
Spatially, the  elements that become intermittent concentrate in a coherent inlet-side corridor, rather than appearing randomly across the domain.
Most of these elements follow the expected monotonic trend (P6, P10, P11 and six of the throats), transitioning from fixed occupancy at the lowest $\mathrm{Ca}$ to intermittent at higher $\mathrm{Ca}$.
This is consistent with rising viscous forcing progressively unblocking capillary-pinned pathways. The clearest example is P11: fixed at $\mathrm{Ca}=3.08\times10^{-6}$, but intermittent from $\mathrm{Ca}=6.82\times10^{-6}$ onwards.

The remaining $\mathrm{Ca}$-sensitive elements deviate from this monotonic pattern. For instance, pore P9 becomes intermittent at $\mathrm{Ca}=6.82\times10^{-6}$ and $1.30\times10^{-5}$ and then re-fixes from $\mathrm{Ca}=1.62\times10^{-5}$ upward; as we show later increasing $\mathrm{Ca}$ leads to a rearrangement of flow pathways and hence some previously intermittent pores can become fixed again, Sec.~\ref{sec:pore_structure}.

\begin{figure}[t]
  \centering
  \includegraphics[width=\columnwidth]{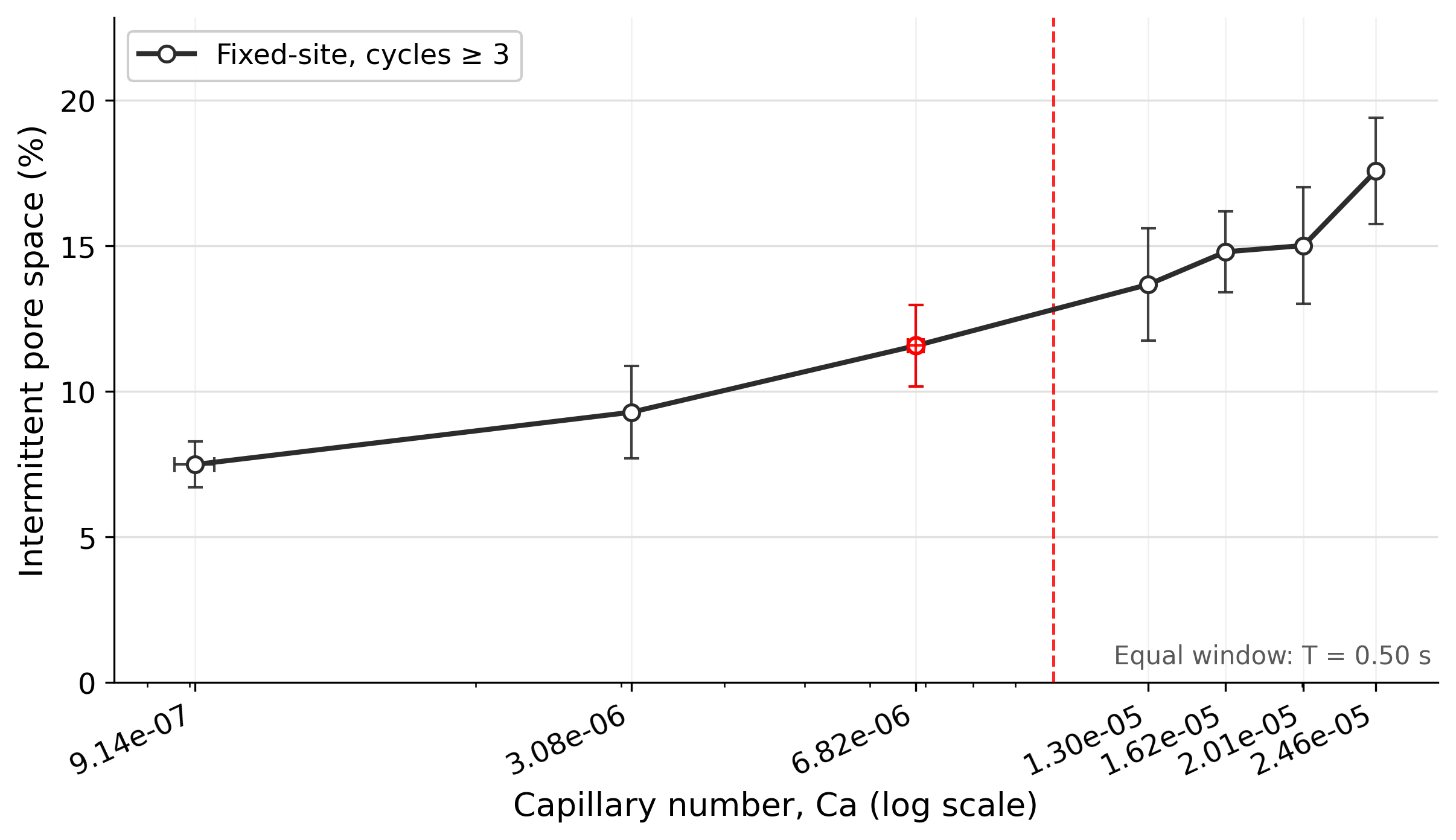}
  \caption{Fixed-site intermittent pore-space fraction as a function of $\mathrm{Ca}$, computed over a common $T = \SI{0.50}{s}$ steady-state window for each case. The vertical dashed red line at $\mathrm{Ca}\approx 10^{-5}$ marks the Darcy-to-intermittent regime transition, and we mark the transition case in red too (Table~\ref{tab:cases}). Horizontal error bars show the uncertainty in $\mathrm{Ca}$ from the variability of inlet flux within the analysed window; vertical error bars combine the binomial counting uncertainty of the intermittent fraction with the standard deviation across equal-window results at $T=$ 0.30, 0.35, and 0.50\,s.}
  \label{fig:interm_vs_ca}
\end{figure}

The same Ca-dependence is reflected in a voxel-level quantification of intermittent activity.
Using the fixed-site local-event classifier (Sec.~\ref{sec:classification}) on a common $T = \SI{0.50}{s}$ steady-state window for each case, we compute the fraction of pore-space voxels classified as intermittent. 
Fig.~\ref{fig:interm_vs_ca} shows that this fraction increases monotonically with capillary number, from \SI{7.5}{\percent} at $\mathrm{Ca}=9.14\times10^{-7}$ to \SI{17.6}{\percent} at $\mathrm{Ca}=2.46\times10^{-5}$. 
The growth rate is itself Ca-dependent: the slope of intermittent fraction against Ca is shallower below the Gao \emph{et al.}~\cite{Gao20} defined regime transition from linear to sub-linear at $\mathrm{Ca}\approx 10^{-5}$ (vertical dashed line), but steeper above it. 
This increased rate of opening-up of new intermittent pore space in the intermittent regime is the pore-scale signature of the macroscopic sub-linear $\nabla P$-Ca scaling reported in Sec.~\ref{sec:macro}: as a larger fraction of the pore space becomes intermittently active, the effective fluid mobility rises and the macroscopic pressure gradient grows sub-linearly with Ca.

\begin{figure}[t]
  \centering
  \includegraphics[width=\columnwidth]{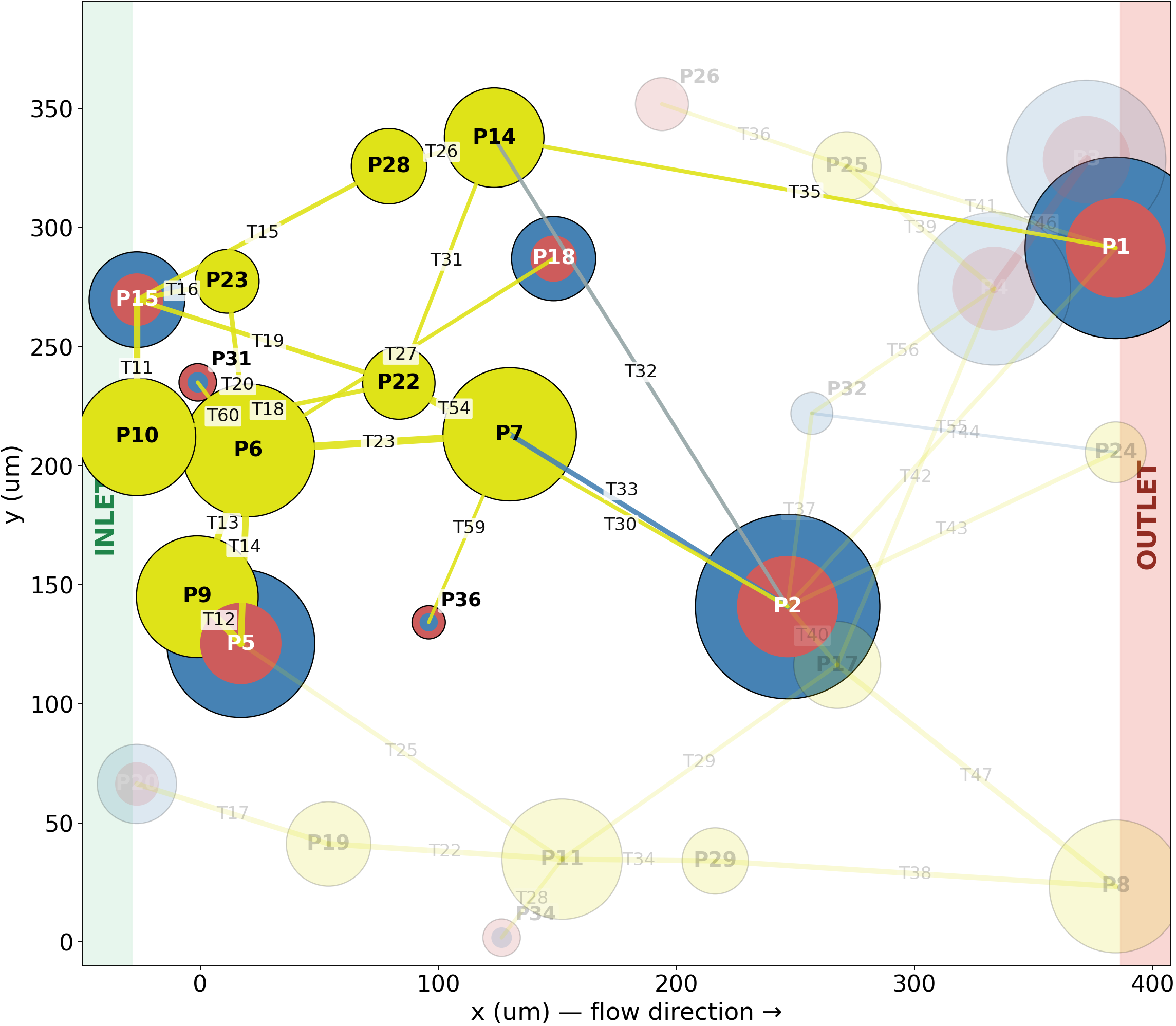}
  \caption{Local network map for the transitional $\mathrm{Ca}\approx 6.82\times10^{-6}$ case, highlighting the intermittent P6-centred cluster within its fixed backbone perimeter. Yellow pores and throats belong to the connected intermittent corridor. Blue and red indicate the wetting and non-wetting phases respectively. Pale elements with low-transparency lie outside the P6-connected intermittent cluster.}
  \label{fig:network_phase_map_p6_focus}
\end{figure}

The physical picture of intermittency at the network scale, motivated by the experimental observations of Spurin~\emph{et al.}~\cite{Spu21}, is the following: the intermittent elements do not appear as isolated unstable sites scattered across the pore space, but form a connected chain of pores and throats (an intermittent conduit) that acts as a bridge between fixed flow pathways.
Fig.~\ref{fig:network_phase_map_p6_focus} isolates this conceptual picture for the $\mathrm{Ca}=6.82\times10^{-6}$ case, showing a contiguous network of intermittent elements centred on pore P6 surrounded by a perimeter of fixed elements that form the backbone on either side.
The cluster was identified by starting from P6 and walking outward through the network graph along intermittent elements, halting at the first fixed element encountered on each branch; the highlighted set is the union of pores and throats mutually reachable from P6 without crossing a fixed conduit, and the surrounding fixed elements are precisely the perimeter at which the walk terminates.
In practice, our $100^3$-voxel subvolume is too small to capture the conceptual picture in full: the intermittent cluster reaches the inlet and outlet boundaries before the surrounding fixed backbone closes around it on those two sides, so the bridge is truncated at the boundaries. The figure nevertheless makes the intermittent passage and the surrounding fixed perimeter directly visible at the network scale, and it identifies the local elements that enter the cluster-scale analysis of Sec.~\ref{sec:p6_cluster}.

Across the entire Ca sweep, we identify two such intermittent conduits coexisting in the present subvolume, each embedded in the same fixed-backbone perimeter (P1, P2, P4, P5, P15, P32 and the connecting throats).
The first is the upper conduit already shown in Fig.~\ref{fig:network_phase_map_p6_focus}, anchored on the stable five-pore set [P6, P7, P10, P22, P23].
The second is a lower conduit, anchored on [P8, P11, P17, P19, P29] and recovered by the same walk starting from P19.
Both are present at every Ca above $\mathrm{Ca}=6.82\times10^{-6}$.

The two conduits show distinct kinds of Ca dependence.
The lower conduit has a discrete activation Ca below which it does not exist as a connected entity: P11 is the gating element, transitioning from fixed to intermittent at $\mathrm{Ca}=6.82\times10^{-6}$ and thereby connecting P19 to P29 to form the conduit. 
Below this Ca, P19 and P29 remain intermittent in isolation; above it, the lower conduit's identity is unchanged across the rest of the sweep.
The upper conduit, by contrast, is present at every Ca with the same five-pore core, while its marginal members (P9, P14) are activated and de-activated non-monotonically (P9 only at the two intermediate Ca; P14 at three of six cases).
In both cases, what changes with rising Ca is not the spatial identity of the conduit's stable core but whether the conduit exists at all (lower conduit) or which marginal pores are recruited around an existing core (upper conduit).
Once formed, the conduit's stable core occupies a Ca-invariant set of pores; rising Ca then primarily modulates how strongly that stable core is activated (Sec.~\ref{sec:pore_structure}, Fig.~\ref{fig:pathway_comparison}b,c) rather than reorganising completely.

\subsection{Macroscopic comparison: pressure gradient as a function of capillary number}
\label{sec:macro}

\begin{figure}[h]
  \centering
  \includegraphics[width=\columnwidth]{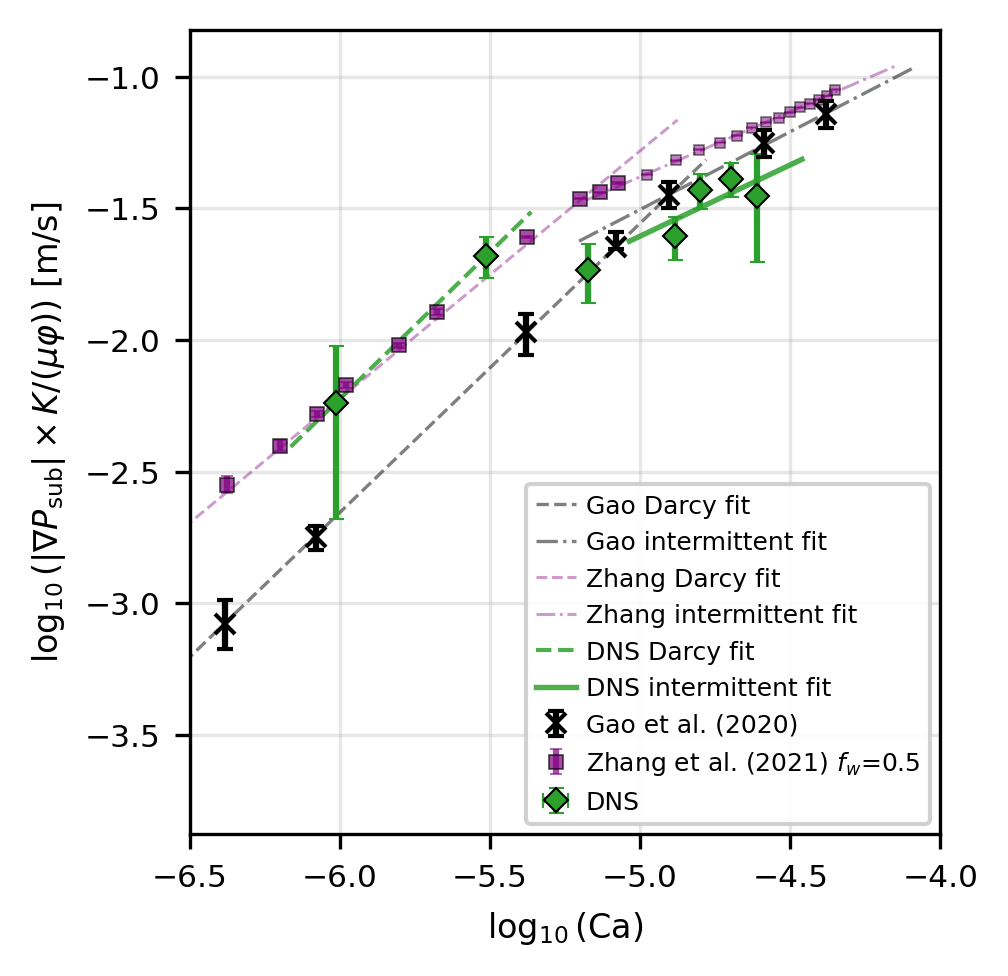}
  \caption{Macroscopic pressure-gradient comparison of the simulation results, DNS, with experiment. With the boundary-condition-affected end regions excluded, both regimes are recovered from the simulation: Darcy slope $1.09$ and intermittent slope $0.54\pm 0.4 $ ($r^2=0.46$), compared with the experimental references of $1.0$ and $0.6$.  Lines are power-law best fits over the regime sets (Darcy: cases 10w10 and 50w50; intermittent: cases 200w200 to 500w500; see Table~\ref{tab:cases}).}
  \label{fig:gradP_Ca}
\end{figure}

We compared the simulation results with the experimental results of Gao \emph{et al.}~\cite{Gao20} and Zhang \emph{et al.}~\cite{Zha21} both of which were also measured on water-wet Bentheimer sandstone samples. Fig.~\ref{fig:gradP_Ca}
shows the pressure gradient normalised by $K/(\mu\phi)$ as a function of $\mathrm{Ca}$ defined in Eq.~(\ref{Eq:Ca}).
The DNS points follow the same overall capillary-number trend as the experiments and lie between the two experimental datasets, indicating comparable scaling but with a modest residual vertical offset. 
The sample size is less than a representative elementary volume, so a close quantitative match with experiment is not expected, as discussed in the Supplementary Material S1.

The recovered sub-linear slope in the intermittent regime is the manifestation of the network-level reorganisation of flow pathways established in Sec.~\ref{sec:network_pathways} (Fig.~\ref{fig:interm_vs_ca}): as a larger fraction of the pore space becomes intermittently active with rising Ca, the effective fluid mobility rises and the macroscopic pressure-gradient-flow-rate relationship is correspondingly sub-linear rather than linear.

\subsection{Temporal organisation of intermittency}
\label{sec:temporal}

We next examine the temporal organisation of the intermittent activity. 
The aim is to identify a characteristic timescale of the per-element connectivity signal and test whether it scales with Ca in the range $\mathrm{Ca}=3.08\times10^{-6}$ to $\mathrm{Ca}=2.46\times10^{-5}$.

\begin{figure}[t]
  \centering
  \includegraphics[width=\columnwidth]{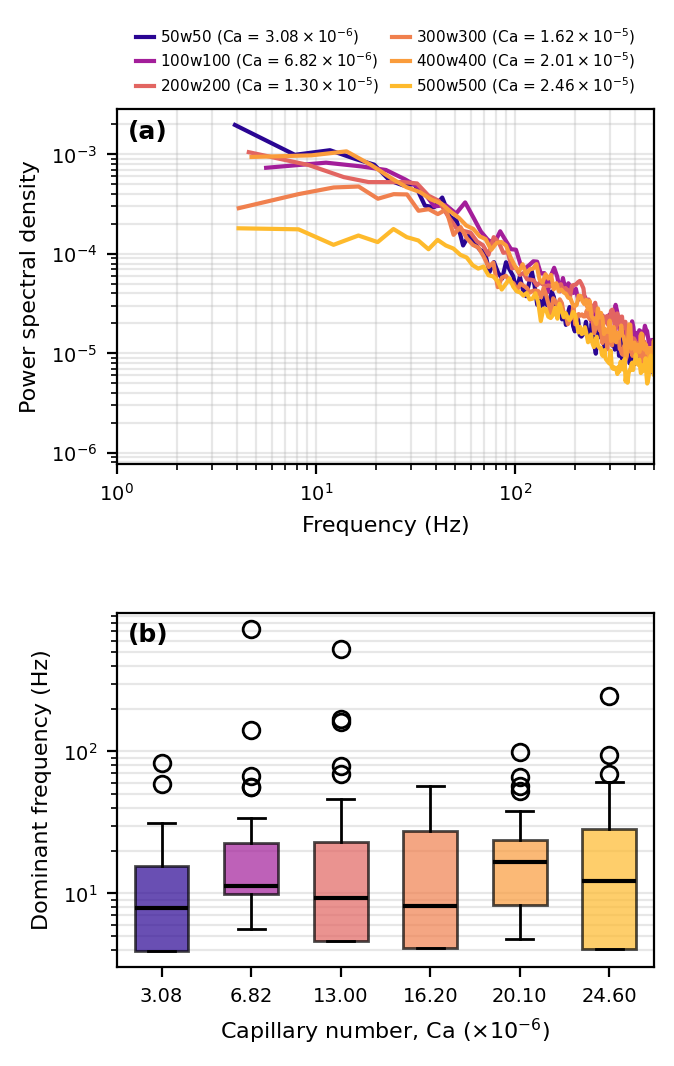}
  \caption{Spectral characterisation of intermittency from the element-based non-wetting spanning signal, using the last \SI{0.75}{s} of the steady-state window for each case. (a)~Median power spectral density, PSD, over all intermittent elements in each case. (b)~Distribution of per-element dominant frequencies. The medians lie in the \SIrange{8}{17}{Hz} band and the interquartile ranges overlap strongly.}
  \label{fig:intermittency_spectral}
\end{figure}

We find that the spectral signature is invariant across the Ca range studied.
Panel~(a) of Fig.~\ref{fig:intermittency_spectral} shows the median power spectral density, PSD, for each case over the common \SI{0.75}{s} window obtained from the element-based connectivity classifier.
All six curves collapse onto a very similar shape: a broad low-frequency plateau followed by a high-frequency decay, with only modest case-to-case variation in overall magnitude.
The shape itself is physically informative.
The absence of any sharp peak shows that the connectivity signal is not periodic: there is no single timescale that a clean disconnection-reconnection cycle would impose.
The signal energy is instead spread broadband across an order-of-magnitude range of frequencies, indicating that the disconnect-reconnect sequence at each element is irregular in time.
The low-frequency plateau corresponds to the slow envelope of these repeated cycles, while the high-frequency decay reflects the increasingly fast sub-cycle interface fluctuations.
That this broadband form is seen for all $\mathrm{Ca}$ implies the temporal fingerprint of intermittency is set by the intrinsic pore-scale capillary-viscous physics rather than by the imposed bulk-flow rate.

Panel~(b) shows the distribution of per-element dominant frequencies.
The medians remain in the \SIrange{8}{17}{Hz} band across all cases, while the interquartile ranges are broad and strongly overlapping. A power-law fit to the median dominant frequency gives only a weak increase with Ca, approximately $f \propto \mathrm{Ca}^{0.2 \pm 0.1}$.
The interquartile ranges of the per-element dominant frequencies overlap strongly across all six cases (their common band is roughly \SIrange{10}{16}{Hz}), so the differences between case medians are small relative to the within-case spread. 
The wide confidence interval reflects two distinct noise sources on each per-case median: each per-element peak pick is quantised at the Welch frequency resolution (Sec.~\ref{sec:spectral_method}, \SIrange{4}{6}{Hz} at this window length, comparable to the median frequencies themselves), and the median is estimated from a finite number of intermittent elements per case ($n_{\mathrm{int}} \approx 40$).
The standard errors associated with $n_{\mathrm{int}}$ for the per-case medians are \SIrange{1.8}{3.5}{Hz}, also comparable to the case-to-case differences themselves.

The dominant frequency is therefore consistent with a flat or only weakly increasing scaling with Ca. 
This is physically significant, since a simple flow-rate scaling (for example, mass conservation with a fixed event volume per disconnection ($f \sim Q/V_{\mathrm{event}}$) or an advective residence time across a pore ($f \sim v/L$)) would predict $f \propto \mathrm{Ca}^1$. 
The observed exponent of $0.2\pm 0.1$ is much smaller, indicating that the characteristic frequency is not set directly by the imposed bulk-flow rate. 
The relevant timescale may be set primarily by pore geometry and local capillary filling–draining dynamics, with Ca acting mainly to recruit different active elements rather than to accelerate the intrinsic oscillation frequency of each element.

At the network scale, this recruitment-of-elements picture is consistent with the energy-balance framework of Zhang~\emph{et al.}~\cite{Zha21}, in which the onset of the nonlinear intermittent regime is predicted by viscous dissipation overtaking the available capillary energy across the network: as Ca rises, more pore-scale sites cross this threshold and become candidates for intermittent activity.

\begin{figure*}[t]
  \centering
  \includegraphics[width=0.92\textwidth]{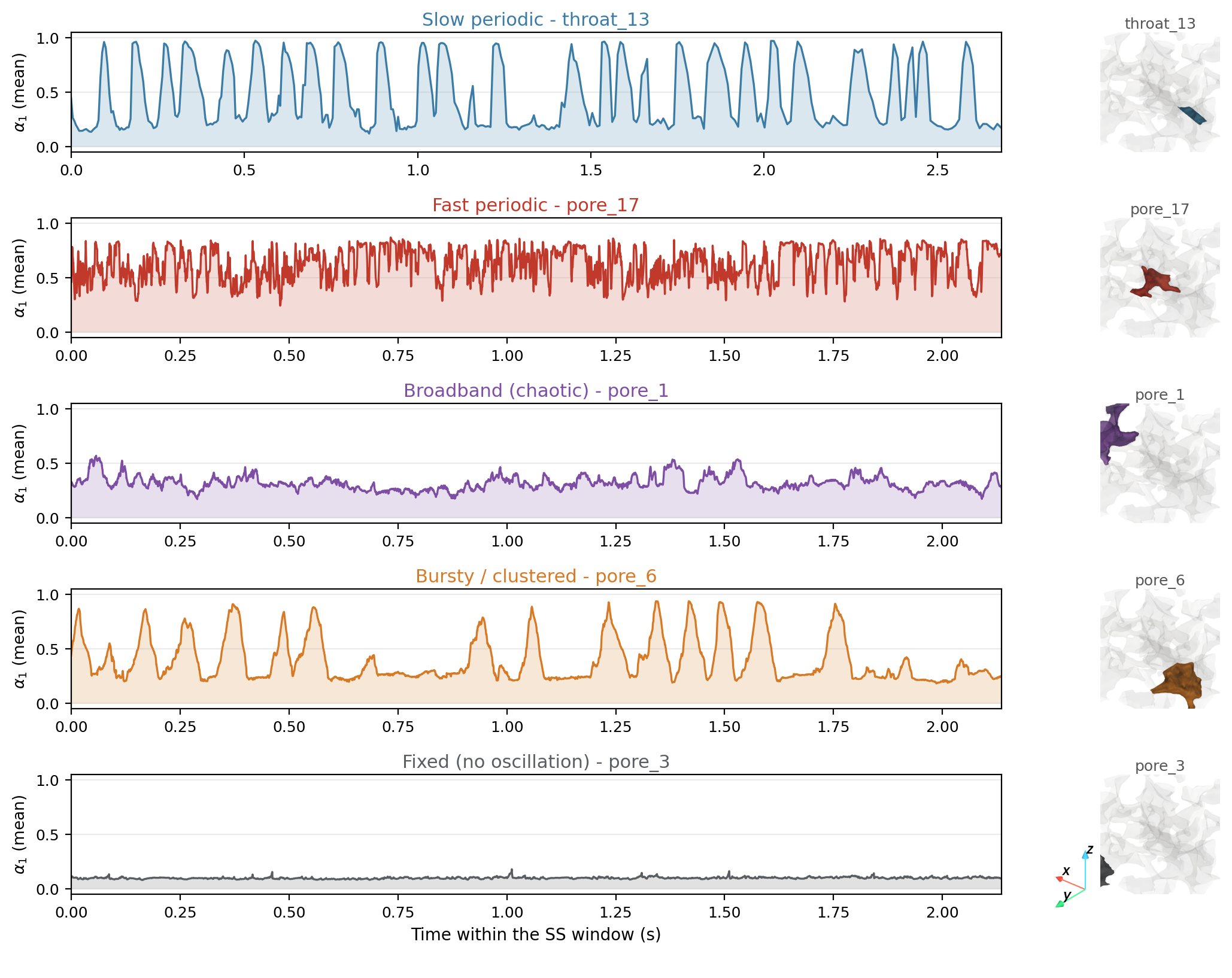}
  \caption{Representative time-domain modes of the element-based intermittency signal for $\mathrm{Ca}=6.82\times10^{-6}$. Left: element-mean $\alpha_1$ signal as a function of time in the steady-state window. Right: 3D position of the corresponding element within the domain. The examples illustrate slow periodic behaviour (T13), faster periodic fluctuations (P17), broadband low-amplitude fluctuations (P1), bursty clustered intermittency (P6), and a fixed non-oscillating element (P3).}
  \label{fig:intermittency_modes}
\end{figure*}

Although the aggregate PSDs are broadly similar across Ca, the time-domain expression of intermittency varies substantially from one element to another.
Fig.~\ref{fig:intermittency_modes} shows representative examples drawn from the element-based signal: a slow periodic throat-scale oscillator, a faster periodic pore-scale signal, a low-amplitude broadband element, a bursty clustered response, and a fixed element that remains non-wetting filled throughout the window.
The accompanying 3D locations show that these modes are distributed across different parts of the domain rather than clustering at a particular element type or spatial region.
This reinforces the interpretation above that the intermittent network contains a heterogeneous population of active elements whose individual temporal signatures differ, even though their aggregate spectral statistics remain broadly Ca-invariant.

\subsection{Intermittency and pore structure}
\label{sec:pore_structure}

Intermittency is shaped by local and network scale structural variables. 
Locally, there is a constriction size dependency: it concentrates wherever the pore space narrows, regardless of whether the narrow region sits inside a throat or within a pore.
At the network scale, the proximity to the hydraulic flow corridor is important: elements along the high-flux inlet-to-outlet route have more intermittent activity, while elements along the geometrically shortest route have less. 

\begin{figure}[t]
  \centering
  \includegraphics[width=\columnwidth]{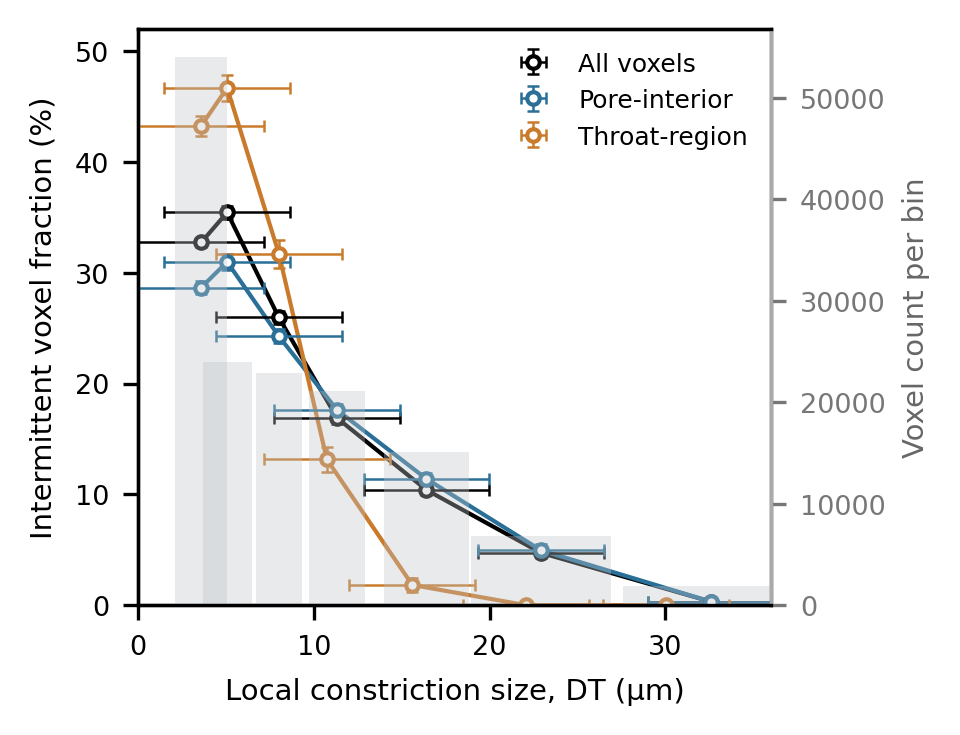}
  \caption{Dependence of intermittency on local constriction size for $\mathrm{Ca}=6.82\times10^{-6}$. Lines show the intermittent voxel fraction binned by local distance transform (DT): black for all pore-space voxels, blue for the pore-interior subset, and orange for the throat-region subset.  Grey bars on the secondary axis show the voxel count per DT bin. Vertical error bars show the 95\% confidence interval from binomial counting uncertainty within each DT bin. Horizontal error bars show a one-voxel geometric localisation uncertainty in DT (${\pm}\SI{3.58}{\micro\metre}$).}  
  \label{fig:constriction_size}
\end{figure}

At the local scale, the distance transform (DT), i.e.\ the distance from each pore-space voxel to the nearest grain surface, is used as a continuous proxy for local constriction size in Fig.~\ref{fig:constriction_size}.
Using this measure, intermittency is strongly concentrated in narrow regions of the pore space: the intermittent fraction is \SIrange{32.7}{35.5}{\percent} for DT below \SI{7}{\micro\metre}, then decreases steadily with increasing DT to \SI{0.31}{\percent} for DT between \SI{29}{\micro\metre} and \SI{40}{\micro\metre}.
The same trend survives when the voxels are separated into pore-interior and throat-region subsets.
Throat-region voxels are more active at the smallest constrictions, reaching \SIrange{43.3}{46.7}{\percent} for DT below \SI{7}{\micro\metre}, but pore-interior voxels also show substantial activity in that same narrow-DT range (\SIrange{28.7}{31.0}{\percent}).
This indicates that intermittency is not confined to extracted throats alone: it is localised wherever the pore space narrows sufficiently, including internal constrictions embedded within pores.

This local-geometry result links naturally to the representative pore-cluster mechanism discussed above.
In Fig.~\ref{fig:p6_flux_balance}, the decisive disconnection does not occur in the larger downstream outlet throat, but at an internal constriction embedded within the connected pore cluster itself. Narrow constrictions are the \emph{preferred locations} of intermittency, but the \emph{activation} of those constrictions is controlled by the connected network dynamics rather than by local geometry alone.

\begin{figure*}[t]
  \centering
  \includegraphics[width=\textwidth]{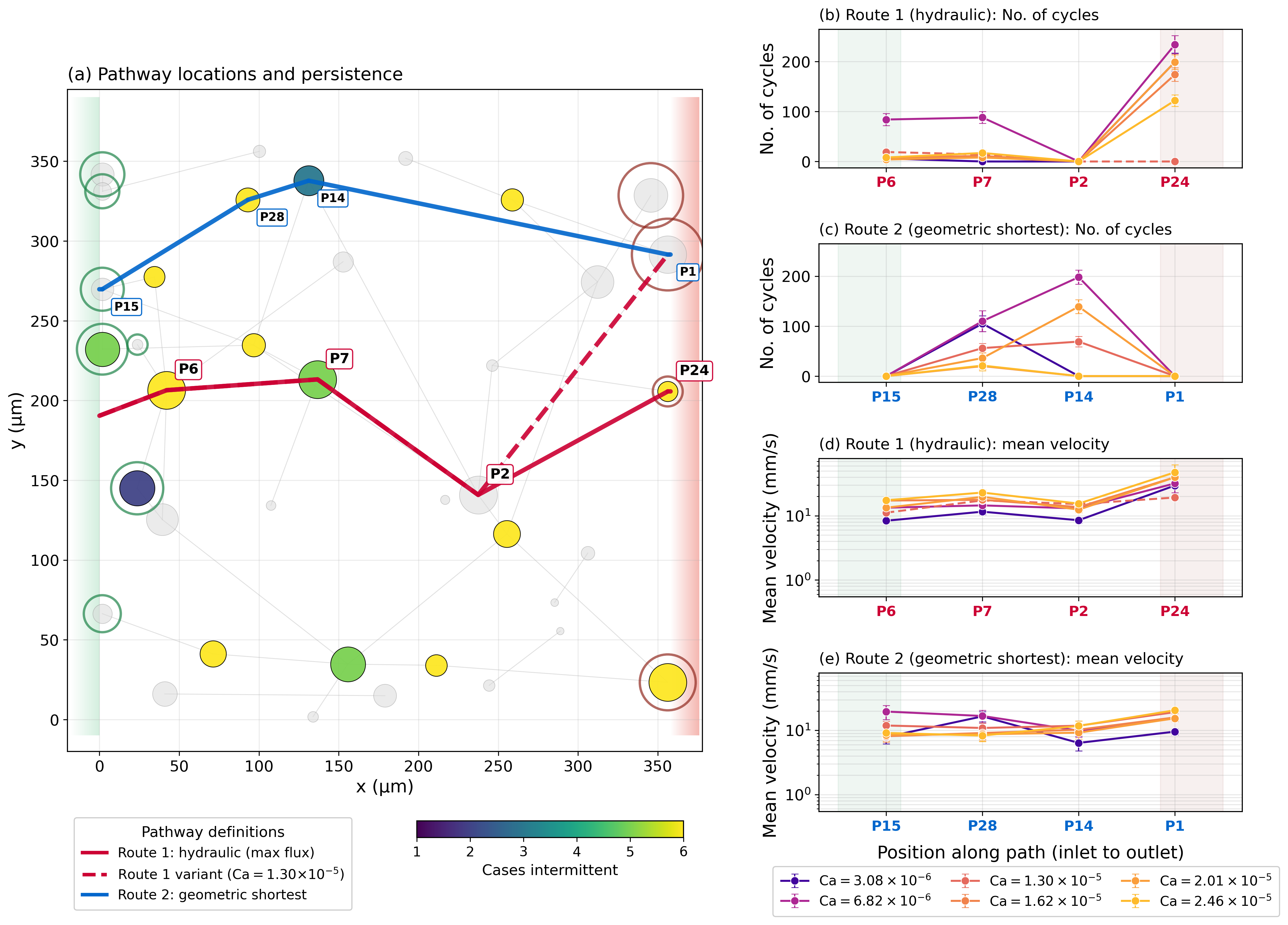} 
  \caption{Two structural pathway definitions and the associated intermittency and flow profiles along each. (a)~Pore-network projection coloured by per-pore intermittency persistence, i.e.\ the number of cases (1 - 6) in which a pore is labelled intermittent under the element-based classifier; pores never intermittent across the sweep are rendered grey. Route~1 (red, solid) is the hydraulic flow-selecting path [P6, P7, P2, P24]; the dashed red overlay shows the single per-Ca variant terminal pore [P6, P7, P2, P1] at $\mathrm{Ca}=1.30\times10^{-5}$. Route~2 (blue, solid) is the geometric shortest inlet-to-outlet path [P15, P28, P14, P1], chosen as the one with smallest physical traversal length (\SI{438}{\micro\metre}). (b),(c)~No.\ of cycles per pore along Route~1 (b) and Route~2 (c), shared y-axis. Error bars combine Poisson $\sqrt{N}$ and the standard deviation of the cycle count re-evaluated over the last 50\,\%, 75\,\% and full SS window (each rescaled to full-window equivalent), added in quadrature. (d),(e)~Mean per-pore velocity along Route~1 (d) and Route~2 (e), log scale, shared y-axis. Error bars combine the spatial sample-mean uncertainty with a placeholder algorithmic uncertainty estimate (5\,\% of the mean) in quadrature. Lines coloured by Ca in all four right-hand panels; the dashed line in panels (b),(d) corresponds to the Route~1 variant terminal pore at $\mathrm{Ca}=1.30\times10^{-5}$.}
  \label{fig:pathway_comparison}
\end{figure*}

At the network scale, intermittency is most prevalent along the hydraulic flow corridor, and not along the geometrically shortest route.
Fig.~\ref{fig:pathway_comparison} compares two inlet-to-outlet routes.
Route~1 is the \emph{hydraulic} (flow-selecting) path, defined as the inlet-to-outlet path that maximises the time-averaged hydraulic flux through the network.
Route~2 is the \emph{geometric shortest} path, defined as the inlet-to-outlet path of minimum total physical traversal length (sum of inscribed throat lengths plus through-pore segments) among all minimum-hop paths through the static pore-network graph.
The two routes are spatially distinct: Route~1 [P6, P7, P2, P24] traces the upper portion of the network and is one of 12 paths tied at the minimum hop count of 3; by physical traversal length it is the 5th-shortest of those 12 (\SI{516}{\micro\metre}). 
Route~2 [P15, P28, P14, P1] is the actual physical-shortest path (\SI{438}{\micro\metre}), running across the centre of the network. 

Route~1's hydraulic identity is largely Ca-invariant, and the disconnection cycles concentrate along it rather than along the geometric route. 
Five of six retained cases share [P6, P7, P2, P24], with the first three pores constant across all retained Ca; only at $\mathrm{Ca}=1.30\times10^{-5}$ does flow redistribute and the path terminate at P1 instead of P24 (dashed in panel a). 
Route~2 has no Ca dependence by construction. 
Along Route~1 (panel b), three of four pores host substantial disconnection activity: P6, P7, and P24 are intermittent in every case, with P2 the single exception (a fixed outlier on the high-flux corridor).
In contrast, along Route~2 (panel c) only the two middle pores P28 and P14 carry cycles in some Ca cases, while the inlet-side P15 and outlet-side P1 are stably fixed.

The mean per-pore velocity profiles (panels d, e) show that their differences are concentrated at the outlet-side pore rather than spread uniformly along the routes. 
Along Route~1 (panel d) the mid-route pores P6, P7, P2 carry velocities in the \SIrange{8}{23}{\milli\meter\per\second} range, while the outlet pore P24 reaches \SIrange{30}{48}{\milli\meter\per\second}. 
Route~2 (panel e) carries comparable mid-route velocities (\SIrange{6}{20}{\milli\meter\per\second}), but its outlet pore P1 caps at \SIrange{10}{21}{\milli\meter\per\second} (roughly half the flux of Route~1's P24), consistent with flow selecting the wide-throat exit at P24 over the geometrically shorter exit at P1. 
Both routes show a shape-preserving profile across Ca: the same pores accelerate and decelerate the fluid in the same order, with only the magnitude shifting upward with Ca.

The comparison shows that network position relative to the hydraulic corridor is a partial predictor of intermittency: 3 of 4 pores on Route~1 are persistently intermittent, well above the network-wide frequency of 10 out of 39 pores. 
The geometric shortest route does not show the same level of intermittency (only 2 of 4 of its pores carry verified intermittent activity).
Neither route, however, captures the full intermittent population of the subvolume: panel~(a) shows a substantial off-route intermittent set scattered around the routes, consistent with the constriction-size analysis above that places preferred-intermittency sites wherever the pore space narrows sufficiently, on or off any privileged route.

\section{Conclusions}
\label{sec:conclusions}

We have presented a systematic direct simulation-based characterisation of intermittency in two-phase flow through a natural porous medium: this is when there is alternating pore-scale occupancy even though capillary forces dominate at the pore scale.
Our main findings are as follows.
\begin{enumerate}
  \item A pore-scale zoom on a representative pore cluster reveals that intermittency consists of repeated, reversible disconnection and re-connection events coupled to strong local pressure redistribution rather than to random interface noise. Disconnection follows a drainage-dominated flux imbalance in which downstream non-wetting phase withdrawal overtakes upstream replenishment before critical snap-off occurs at a narrow supply throat.
  \item At the network scale, the intermittent pore set decomposes into two coexisting connected conduits embedded in a stable surrounding backbone of fixed flow pathways, reproducing the conduit picture of Spurin~\emph{et al.}~\cite{Spu21} across the intermittent range studied. Increasing Ca leads to more of the pore space becoming intermittent, and the enhanced degree of intermittency leads to the transition from the linear to the sub-linear regime.
  \item The simulations reproduce both the linear Darcy slope and the sub-linear intermittent $\nabla P$-Ca slope observed experimentally by Gao \emph{et al.}~\cite{Gao20} and Zhang \emph{et al.}~\cite{Zha21}.
  \item The temporal organisation of the intermittent activity has no characteristic timescale: the per-element connectivity signal is broadband and Ca-invariant within regression uncertainty ($f \propto \mathrm{Ca}^{0.19 \pm 0.13}$), indicating that the temporal fingerprint of intermittency is set by intrinsic pore-scale capillary-viscous physics rather than by the imposed bulk-flow rate.
  \item Intermittent activity is concentrated at narrow constrictions, including constrictions embedded within pores. Network position relative to the hydraulic flow corridor partially predicts intermittency at the per-pore level, although a substantial intermittent population persists off-route. Local geometric susceptibility is therefore necessary but not sufficient: activation of intermittent sites is controlled by network-coupled pressure and flux redistribution.
\end{enumerate}

These results support a physical picture in which intermittency arises from the repeated exploration of metastable interface configurations within a connected pore network.
Capillary barriers select the susceptible constrictions, viscous forcing recruits an increasing fraction of those constrictions as Ca rises, and network-scale pressure redistribution couples the resulting local events into a macroscopic intermittent flow regime.

Future work could compare simulation predictions with time-resolved experiments on a pore-by-pore basis, and explore different geometries and wettability. 
Furthermore, simulations on a larger domain may provide more robust estimates of macroscopic properties.

\begin{acknowledgments}
AK is grateful for a Janet Watson scholarship from the Department of Earth Science and Engineering that funds her PhD. 
The authors are also grateful to the Imperial College Research Computing Service for the computational resources and support, including access to the CX3 Phase 2 cluster~\cite{ImperialHPC}.
\end{acknowledgments}

\appendix
\section{Supplemental video: P6 cluster intermittency event}
\label{sec:p6_video}

The supplemental video (file: \texttt{P6\_Disconnection\_Event.mp4}) shows the same eight panels as Fig.~\ref{fig:p6_overview} of the main text, animated over the full duration of a representative disconnection-reconnection cycle in the P6 cluster, for the transitional $\mathrm{Ca}\approx 6.82\times10^{-6}$ case (event D5). The time-resolved sequence captures (i) the build-up of the drainage-dominated flux imbalance across the cluster, (ii) the critical snap-off event at T60, (iii) the advection of the detached non-wetting fragment through P6 toward the outlet throat T23, and (iv) the subsequent Haines-jump reconnection that restores cluster connectivity. The viscous-pressure heatmaps in panels~(b),(c),(h) and the velocity-magnitude heatmaps in panels~(f),(g) evolve continuously throughout the cycle, allowing the static snapshot of Fig.~\ref{fig:p6_overview} to be placed in its dynamic context.

\bibliography{references}

\end{document}


\title{Supplemental Material:\\
Intermittent two-phase flow in porous media:\\
insights from pore-scale direct numerical simulation}

\author{Alexandra Karabasova}
\affiliation{Department of Earth Science and Engineering, Imperial College London}

\author{Sajjad Foroughi}
\affiliation{Department of Earth Science and Engineering, Imperial College London}

\author{Martin J. Blunt }
\affiliation{Department of Earth Science and Engineering, Imperial College London}

\author{Branko Bijeljic }
\affiliation{Department of Earth Science and Engineering, Imperial College London}

\date{\today}

\maketitle

\section{Pore geometry and representative elementary volume}
\label{sec:rev}

The simulation domain is a subvolume extracted from a micro-CT scan of Bentheimer sandstone at a resolution of \SI{3.58}{\micro\metre\per voxel}.
The micro-CT images are accessible on the Imperial College Pore-Scale Modelling and Imaging website~\cite{Imp26ct}.

To quantify the representative elementary volume (REV), we performed a systematic convergence analysis of pore-network statistics extracted at subvolume sizes ranging from $100^3$ to $999^3$~voxels using the maximal inscribed sphere algorithm~\cite{Rae17} as implemented in \texttt{poreXtractor}~\cite{Imp26px}. This software provides a topological analysis of the pore space, dividing it into wide regions, pores, connected by narrower restrictions, throats.

\begin{figure*}[!ht]
  \centering
  \makebox[\textwidth][c]{\includegraphics[width=1.08\textwidth]{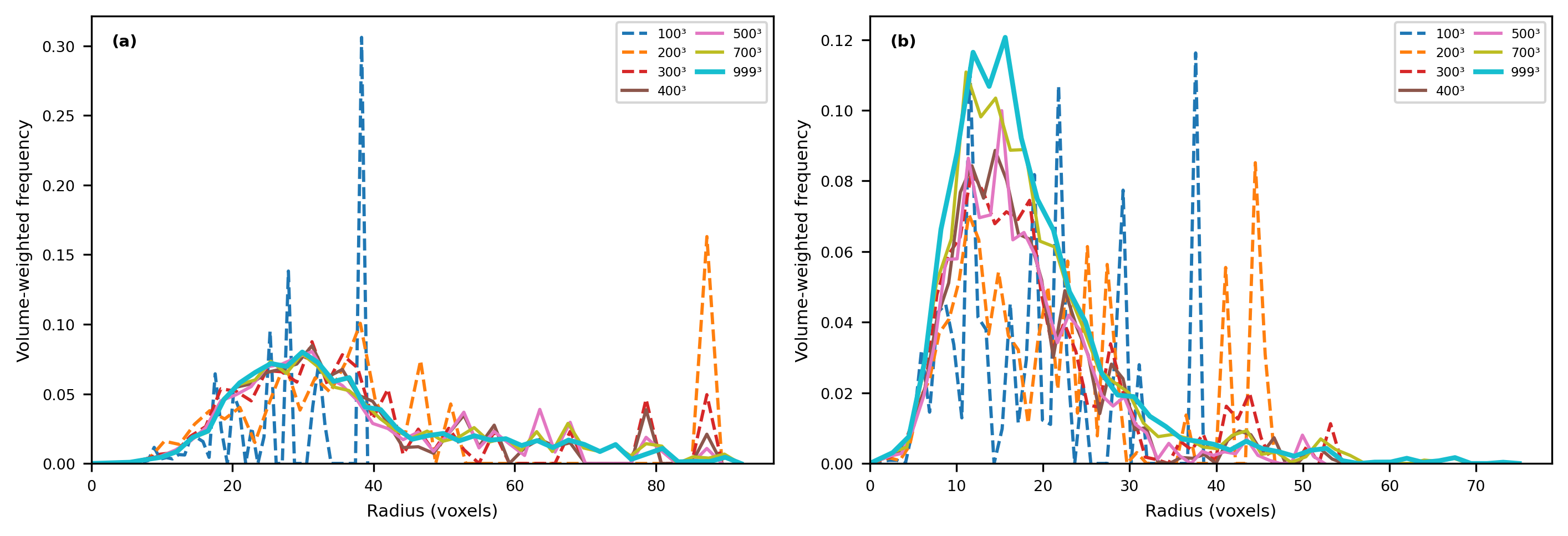}}
  \caption{Volume-weighted PDFs of pore radius (a) and throat radius (b) at different subvolume sizes. Dashed lines denote small subvolumes ($100^3$-$300^3$); solid lines denote larger subvolumes ($400^3$-$999^3$).}
  \label{fig:rev_pdf}
\end{figure*}

Fig.~\ref{fig:rev_pdf} compares the volume-weighted pore and throat radius probability density functions (PDFs) at different subvolume sizes.
At small sizes ($100^3$, $200^3$, $300^3$; dashed lines), the distributions are spiky and erratic, while at $400^3$ and above (solid lines), the PDFs collapse onto a stable reference shape.

\begin{figure}[!t]
  \centering
  \includegraphics[width=\columnwidth]{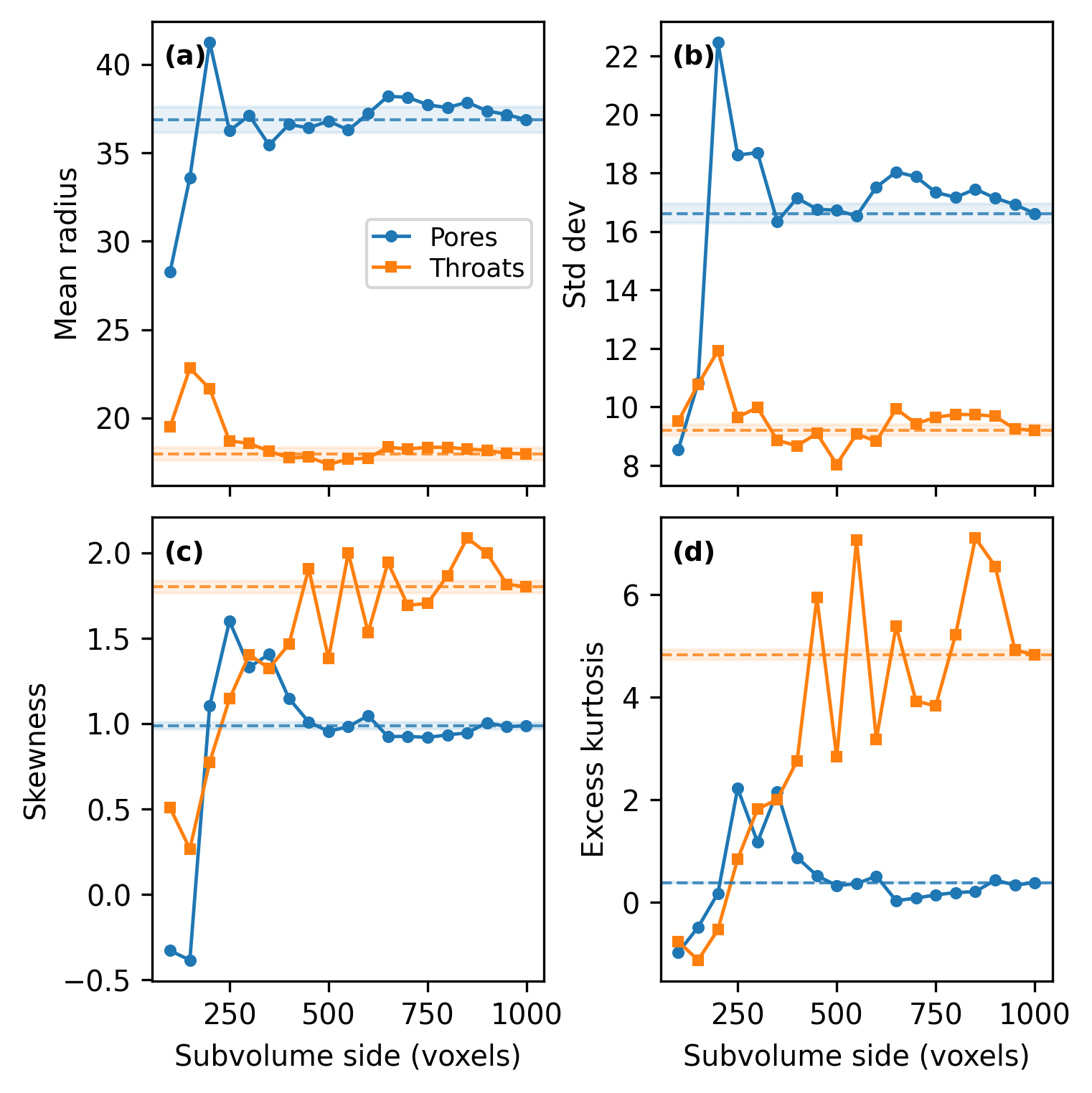}
  \caption{Convergence of distribution moments with subvolume size: mean radius (a), standard deviation (b), skewness (c), and excess kurtosis (d). In each panel, pore and throat statistics are shown together as functions of subvolume side length. Shaded bands denote $\pm 2\%$ of the reference values at $999^3$.}
  \label{fig:rev_moments}
\end{figure}

Fig.~\ref{fig:rev_moments} shows the convergence of the first four distribution moments (mean, standard deviation, skewness, kurtosis) with subvolume size for both pores (in blue) and throats (in orange).
The shaded band denotes $\pm 2\%$ of the reference value (computed for $999^3$).
The pore mean and standard deviation stabilise by approximately $350^3$~voxels, while throat moments require ${\sim}400^3$~voxels.
Higher-order moments (skewness, kurtosis) remain noisy throughout, consistent with the inherent sensitivity of these statistics to rare large pores. 

Based on this analysis, the minimum REV for pore-throat network statistics is approximately $400^3$~voxels.
The maximum domain size currently feasible for time-resolved two-phase DNS is ${\sim}100^3$~voxels (limited by computational cost), which is below the REV.
We note for context that the lattice Boltzmann study of Akai \emph{et al.}~\cite{Aka20}\ used a domain of $288^3$~voxels, and the DNS of Raeini \emph{et al.}~\cite{Rae14}\ used $400 \times 256 \times 256$ but at very short timescales insufficient to reach steady state.
Our $100^3$ subvolume thus represents a trade-off between resolving steady-state intermittent dynamics over a $\sim 2-4s$ range, and approaching a statistically representative geometry. The geometric porosity of the subvolume used for subsequent simulation is $\phi = 0.215$ (214\,756 pore voxels out of $10^6$), while the effective (flow-accessible) porosity is $\phi_\mathrm{eff} = 0.158$.
Both values lie below the bulk Bentheimer porosity (${\approx}0.23$), consistent with the REV analysis in Fig.~\ref{fig:rev_moments} and indicating that this subvolume samples a locally tighter-than-average region of the rock.
The main geometric uncertainty arises from the binary segmentation of pore space and grain space: a ${\pm}1$-voxel variation in the threshold changes the total pore volume by about ${\pm}2\%$, while pore and throat radii from the maximal inscribed sphere algorithm inherit a voxel-scale discretisation uncertainty of order ${\pm}0.5$~voxel (${\pm}\SI{1.8}{\micro\metre}$). This uncertainty is proportionally largest for the smallest throats.

\section{Runtime summary}
\label{sec:runtime_summary}

Table~\ref{tab:runtime_summary} summarises the accumulated runtime of the retained flowing cases used in the final analysis.
We retain the weakly converged 10w10 case in Fig.~\ref{fig:interm_vs_ca} and Fig.~\ref{fig:gradP_Ca}, but exclude it from the spectral and pathway analyses, where its short steady-state window does not compare fairly with the other six cases.

\begin{table*}[t]
\caption{Runtime summary for the cases. Wall-clock time is the total elapsed real time accumulated across the simulation. Core-hours are computed for a single 45-core node. $^{\dagger}$10w10 had reached only a weakly steady state by the end of the simulated time (Table~\ref{tab:ss_windows}).}
\label{tab:runtime_summary}
\renewcommand{\arraystretch}{1.15}
\begin{ruledtabular}
\begin{tabular}{lcccc}
Case & Ca & Wall-clock (h) & Core-hours & Final simulated time (s) \\
\colrule
10w10$^{\dagger}$ & $9.14 \times 10^{-7}$ & 1202.6 & 54\,117 & 5.346 \\
50w50   & $3.08 \times 10^{-6}$ & 1238.6 & 55\,737 & 4.396 \\
100w100 & $6.82 \times 10^{-6}$ &  781.1 & 35\,150 & 2.684 \\
200w200 & $1.30 \times 10^{-5}$ &  433.1 & 19\,489 & 1.558 \\
300w300 & $1.62 \times 10^{-5}$ &  419.2 & 18\,864 & 1.806 \\
400w400 & $2.01 \times 10^{-5}$ &  453.4 & 20\,401 & 1.787 \\
500w500 & $2.46 \times 10^{-5}$ &  314.2 & 14\,139 & 1.363 \\
\end{tabular}
\end{ruledtabular}
\end{table*}

\section{Steady-state window detection}
\label{sec:ss_windows}

The steady-state (SS) window for each case is identified from the domain-averaged wetting-phase saturation $\langle S_1\rangle(t)$ using an adaptive median-based detector. 
A sliding window was swept backward from the end of the signal; at each position, two criteria are evaluated on a median-smoothed saturation trace: (i)~the local linear trend must be below a threshold slope, and (ii)~the median absolute deviation (MAD) of the raw signal, normalised by the local median, must remain below \SI{5}{\percent}.
Both thresholds adapt to the signal character: in the tail-focused part of the signal, where the detector searches the last \SI{80}{\percent} of the simulation time, slope and variability criteria are relaxed when the fluctuations are oscillation-dominated rather than drift-dominated (oscillation ratio ${>}0.85$).
This allowed long quasi-steady windows to be retained even when the saturation continued to oscillate about a stable mean, which is a feature of intermittent steady state, rather than transience.
The SS window is taken as the longest continuous region near the end of the simulation that satisfies these criteria.

Table~\ref{tab:ss_windows} summarises the windows obtained for each simulated case.
For the lowest-Ca flowing case, 10w10, the total run reached only \SI{5.35}{s} of simulated time after approximately \SI{1203}{h} of accumulated wall-clock time. 
Although the SS detector identifies a locally quasi-steady window in the signal (Table~\ref{tab:ss_windows}), the global saturation continues to drift after that window and has not reached a steady value by the end of the run; the case is therefore classified as only weakly steady.

\begin{table*}[t]
\caption{Steady-state windows for all simulated cases, detected from the domain-averaged wetting-phase saturation $\langle S_1 \rangle(t)$. ``Strict'' indicates that both the local trend and the MAD criteria are satisfied within the chosen window. $^{\dagger}$The detector identified a locally quasi-steady window for 10w10, but the global saturation signal had not reached a steady value by the end of the run; this case is therefore labelled ``weak'' and is excluded from some quantitative analyses (see Sec.~\ref{sec:runtime_summary}).}
\label{tab:ss_windows}
\renewcommand{\arraystretch}{1.15}
\begin{ruledtabular}
\begin{tabular}{lccccc}
Case & Ca & $t_\mathrm{start}$ (s) & $t_\mathrm{end}$ (s) & Duration (s) & Reason \\
\colrule
10w10$^{\dagger}$   & $9.14 \times 10^{-7}$   & 2.625 & 4.270 & 1.645 & weak \\
50w50   & $3.08 \times 10^{-6}$   & 2.109 & 3.653 & 1.544 & strict \\
100w100 & $6.82 \times 10^{-6}$   & 1.607 & 2.685 & 1.078 & strict \\
200w200 & $1.30 \times 10^{-5}$   & 1.228 & 1.806 & 0.578 & strict \\
300w300 & $1.62 \times 10^{-5}$   & 1.044 & 1.730 & 0.686 & strict \\
400w400 & $2.01 \times 10^{-5}$   & 1.072 & 1.788 & 0.716 & strict \\
500w500 & $2.46 \times 10^{-5}$   & 0.794 & 1.363 & 0.569 & strict\\
\end{tabular}
\end{ruledtabular}
\end{table*}

The use of the median (rather than the mean) for both the trend and variability metrics makes the detection robust to isolated capillary events.
For reporting macroscopic quantities such as the pressure gradient (Sec.~\ref{sec:gradP_method}), a trimmed mean over the SS window was used instead, as it provides a more physically representative estimate of the time-averaged flow quantities while still reducing the influence of rare extreme events.

\begin{figure*}[t]
  \centering
  \includegraphics[width=\textwidth]{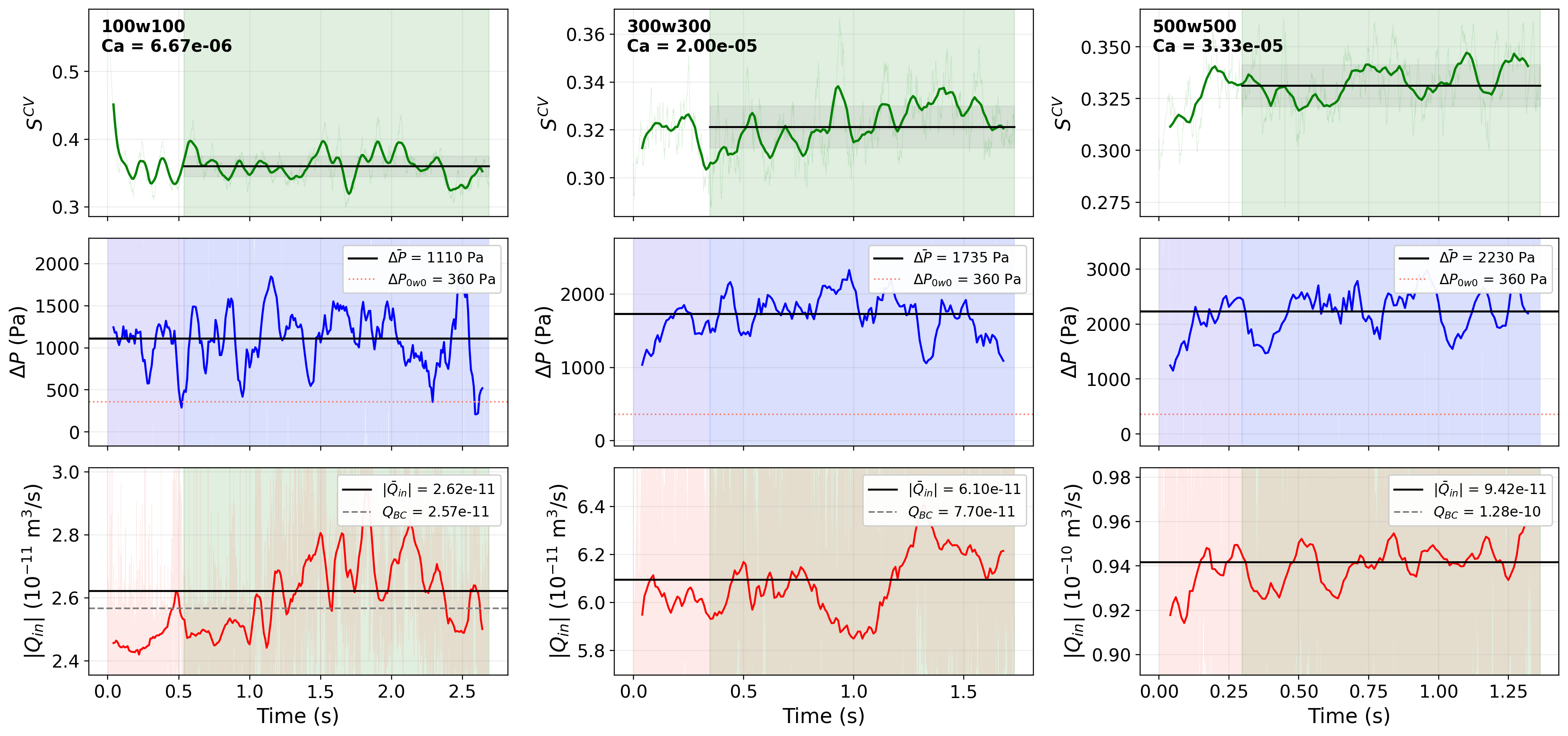}
  \caption{Time series of the three quantities used for SS-window detection and downstream macroscopic averaging: the domain-averaged wetting-phase saturation $\langle S_1 \rangle(t)$ (top row), the boundary pressure drop $\Delta P$ (middle row), and the realised inlet volumetric flow $|Q_{\mathrm{in}}|$ (bottom row). Three representative cases are shown: 100w100 (Ca $=6.82\times10^{-6}$), 300w300 (Ca $=1.62\times10^{-5}$) and 500w500 (Ca $=2.46\times10^{-5}$). In each panel the shaded region marks the SS window selected by the adaptive detector, and the horizontal black line is the within-window mean.}
  \label{fig:gradP_convergence}
\end{figure*}

Fig.~\ref{fig:gradP_convergence} confirms that the SS window selected by the detector captures intervals where the wetting-phase saturation, the boundary pressure drop, and the realised inlet flux all fluctuate around well-defined means, so the time averages reported in the macroscopic comparison (Fig.~\ref{fig:gradP_Ca}) are extracted from a genuinely steady interval rather than from a transient.

\section{Intermittency classifier methodology}
\label{sec:classifier_details}

This section gives the algorithm details and parameter choices for the fixed-site local-event classifier introduced in Sec.~\ref{sec:classification}. The simpler element-based connectivity classifier is fully specified inline in Sec.~\ref{sec:element_classifier}.

\subsection{Details of the fixed-site event-based detector}

The detector operates on pairs of consecutive timesteps within the SS window. 
At each timestep, the pore space was first binarised at $\alpha_1 = 0.5$, with wetting voxels defined by $\alpha_1 \geq 0.5$ and non-wetting voxels by $\alpha_1 < 0.5$.
This gives a binary 3D mask of the instantaneous non-wetting occupancy.

Next, connected-component labelling was applied to the non-wetting mask using 6-connectivity.
Here 6-connectivity means that two non-wetting voxels belong to the same cluster only if they share a face, i.e.\ they are adjacent in one of the six orthogonal directions $(\pm x,\pm y,\pm z)$.
Touching only at an edge or a corner does not count as connected.
The labelling algorithm walked outward through face-adjacent non-wetting voxels and assigned a common cluster ID to every reachable voxel, so that each distinct non-wetting ganglion at that timestep received its own label.

The event-detection step then compared the current timestep with the previous one.
All voxels whose phase label had changed between the two timesteps were identified and grouped into 6-connected flipped clusters; clusters smaller than 5 voxels were discarded as interface noise.
Each retained flipped cluster was treated as a candidate local event.
For each such cluster, a local neighbourhood around the cluster centroid was extracted and a 6-connected flood fill was applied to the non-wetting phase before and after the phase flip to determine how the local non-wetting connectivity changed.
If one non-wetting cluster became two, the event was classified as a snap-off.
If two previously separate non-wetting clusters became one, the event was classified as a drainage coalescence or re-connection event.
If the local non-wetting component count was unchanged, the cluster was discarded.
This local topological test was designed to retain genuine disconnection (imbibition by snap-off) or re-connection (drainage) events while rejecting false positives caused by interface translation, layer swelling, or local occupancy fluctuations that alter phase distribution without fully cutting off or reconnecting a continuous phase pathway.

After the event type had been established, per-voxel counters were updated for the voxels belonging to the validated flipped cluster.
Over the full steady-state window this produced, for each pore voxel, a total event count, a timestamp-indexed record of which timesteps the voxel participated in an event, and a timestamped event log at the global level for the detected snap-off and coalescence events.

Because some recurrent snap-off events migrate slightly in space from cycle to cycle, the fixed-site criterion can miss physically intermittent regions whose break location shifts over time. 
We therefore also computed a \emph{migrating-site} version of the same classifier. 
The sensitivity test revealed that it is consistent with the fixed-site classifier, and is not shown.

\subsection{Cycle-counting and noise filters}

For each voxel, the phase time series was filtered so that each half-phase had to persist for $\geq \SI{1}{ms}$; any transitions separated by shorter intervals were treated as VOF interface noise and not counted towards cycles.
We then counted full event/no-event/event cycles, and classified a voxel as intermittent only if it underwent at least three full cycles.
As a final noise-removal step, any spatially isolated cluster of fewer than 10 classified-intermittent voxels was discarded as residual noise. 
This complements the per-event $\geq 5$ voxel filter (which rejected noise event-by-event) by catching small isolated specks that accumulated just enough cycles to cross the threshold.

\subsection{Equal-window cross-case comparison}

Because the available steady-state windows differ between cases, intermittency cannot be compared fairly using each case's full analysed duration. All cross-case intermittency comparisons therefore use equal physical observation windows taken from the end of the validated SS interval for each case. The intermittency detector is rerun for a range of window lengths $T$, and the resulting intermittent fraction is checked for monotonicity with Ca. The current fixed-site metric gives a strictly monotonic increase across the retained cases for $T \geq \SI{0.50}{s}$. At shorter windows the trend is broadly increasing but with a residual 300w300/400w400 inversion that disappears once $T$ reaches half a second; this is consistent with the inversion being a short-window noise artefact rather than a feature of the underlying intermittent dynamics. This equal-window approach removes the bias that would otherwise arise because longer steady-state windows contain more opportunities to observe rare topology changes.

\section{Sub-box analysis window: boundary contamination and choice of $L_{\mathrm{sub}}$}
\label{sec:subbox_motivation}

\begin{figure*}[t]
  \centering
  \includegraphics[width=\textwidth]{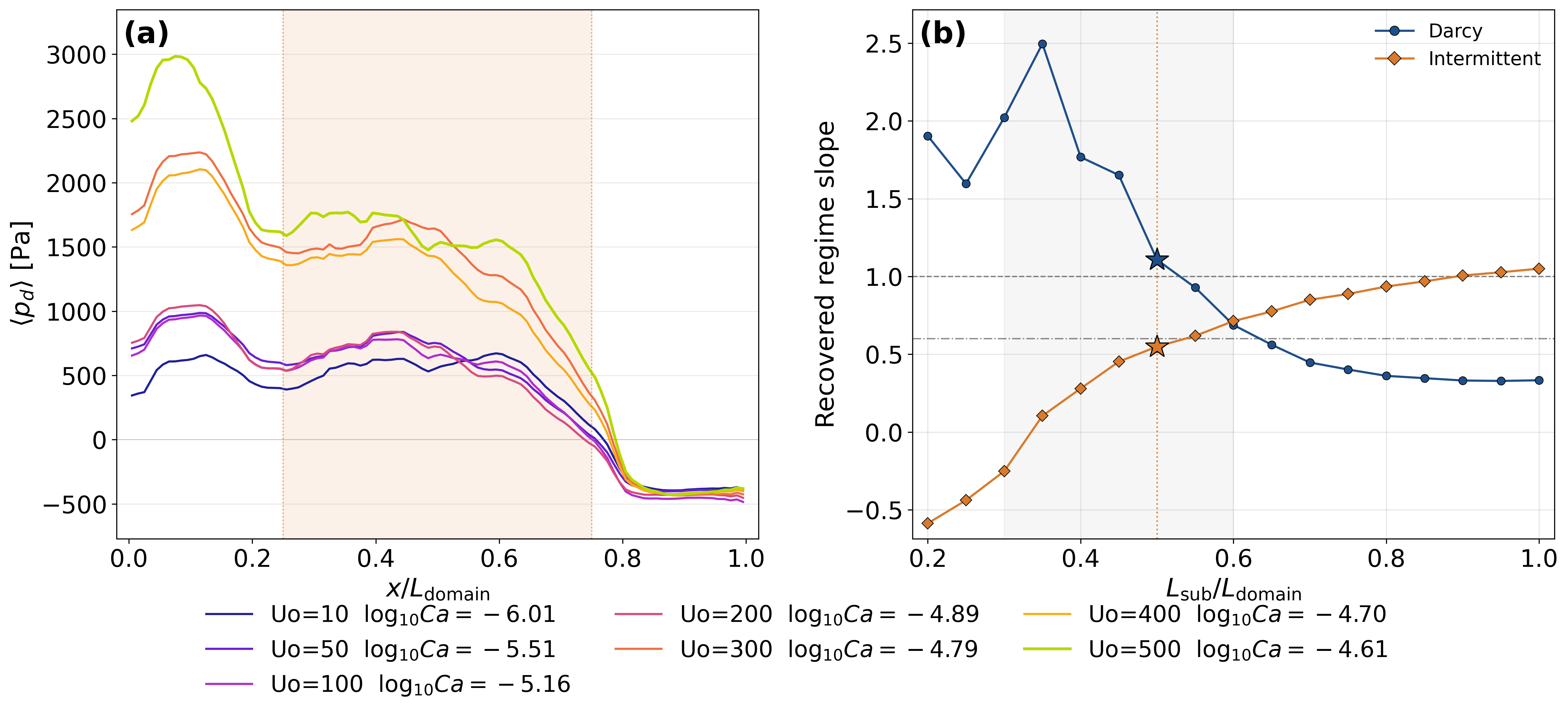}
  \caption{Motivation for the centred sub-box analysis window. Panel~(a) shows the time-averaged streamwise dynamic-pressure profile $\langle p_d\rangle(x)$ for all cases (one curve per $U_o$, coloured by Ca). The lightly shaded central band marks the chosen analysis window, $L_{\mathrm{sub}}/L_{\mathrm{domain}} = 0.50$. Panel~(b) shows the recovered Darcy-regime slope (blue; Uo~$=10, 50$; $n=2$) and intermittent-regime slope (orange; Uo~$=200$ - $500$; $n=4$) of $\nabla P_{\mathrm{sub}}$ versus Ca, as a function of $L_{\mathrm{sub}}/L_{\mathrm{domain}}$. Horizontal reference lines mark the experimental slopes from Gao~\emph{et al.}~\cite{Gao20} ($1.0$ Darcy, $0.6$ intermittent).}
  \label{fig:motivation}
\end{figure*}

The choice of the boundary box for the calculation of pressure gradient, $L_{\mathrm{sub}}$, is data-driven.
Fig.~\ref{fig:motivation}(b) shows that the recovered Darcy slope (fit over the linear-regime cases) rises monotonically as $L_{\mathrm{sub}}/L_{\mathrm{domain}}$ shrinks, while the intermittent slope (fit over the sub-linear-regime cases) is approximately flat across $L_{\mathrm{sub}}/L_{\mathrm{domain}} \in [0.30, 0.60]$.
The two slopes simultaneously match the Gao {\it et al.} reference values at $L_{\mathrm{sub}}/L_{\mathrm{domain}} = 0.50$.
We therefore adopt $L_{\mathrm{sub}}/L_{\mathrm{domain}} = 0.50$ (that is, $25\%$ excluded from each side) as the analysis window throughout.
This choice is large enough to fully exclude the inlet-peak ($x/L < 0.15$) and outlet-relaxation ($x/L > 0.90$) zones, but small enough that the linear regression remains within the bulk plateau of the $\langle p_d\rangle(x)$ profile.
As a separate implementation check, the boundary pressure drop was also compared against a manual area-weighted patch average. The difference was small (about 3.4\%) and was therefore retained only as a systematic uncertainty in the main-text error bars rather than as a separate methodological discussion there.

\bibliography{references}